\documentclass[12pt]{iopart}

\usepackage[letterpaper, margin=1in]{geometry}

\usepackage{xcolor}%
\usepackage{natbib}
\usepackage{graphicx}%
\usepackage{amsmath,amssymb,amsfonts}%
\usepackage{algorithm}
\usepackage{algorithmic}
\usepackage[]{hyperref}
\graphicspath{{Figures/}}
\begin{document}

\title{A fast, matrix-based method to perform omnidirectional pressure integration}

\author{Fernando Zigunov \& John J. Charonko}

\address{Los Alamos National Laboratory, Physics Division, Los Alamos, USA}
\ead{fzigunov@gmail.com}
\ead{john.charonko@lanl.gov}
\vspace{10pt}
\begin{indented}
\item[]August 2023
\end{indented}

\begin{abstract}
Experimentally-measured pressure fields play an important role in understanding many fluid dynamics problems.
Unfortunately, pressure fields are difficult to measure directly with non-invasive, spatially resolved diagnostics, and calculations of pressure from velocity have proven sensitive to error in the data. Omnidirectional line integration methods are usually more accurate and robust to these effects as compared to implicit Poisson equations, but have seen slower uptake due to the higher computational and memory costs, particularly in 3D domains.  
This paper demonstrates how omnidirectional line integration approaches can be converted to a matrix inversion problem.
This novel formulation uses an iterative approach so that the boundary conditions are updated each step, preserving the convergence behavior of omnidirectional schemes while also keeping the computational efficiency of Poisson solvers. This method is implemented in Matlab and also as a GPU-accelerated code in CUDA-C++. The behavior of the new method is demonstrated on 2D and 3D synthetic and experimental data. Three-dimensional grid sizes of up to 125 million grid points are tractable with this method, opening exciting opportunities to perform volumetric pressure field estimation from 3D PIV measurements.
\end{abstract}

\section{Introduction}

The hydrodynamic pressure field plays an important role in the dynamics of the flow in many fluid mechanics problems. 
In particular, the correlation of the pressure fluctuations with the velocity field significantly influences the transport and evolution of turbulence.
Additionally, detecting and quantifying pressure wave sources in the flow is an important part of acoustics and efforts to minimize noise sources, while the interaction of the pressure field with solid bodies such as planes and bridges plays a role in the design of engineered structures.
However, the measurement of these fluctuations has historically been difficult without resorting to the use of invasive pressure probes or surface-mounted sensors or ports, and the measurement of pressure fluctuations and their gradients throughout a spatially-resolved field has been impractical.  
In particular, correlated velocity and pressure measurements needed to explore turbulence theory and models for computational simulations of flows have been lacking.  
To overcome this, researchers have turned to calculation of pressure data from experimentally-measured velocity fields \citep{van_oudheusden_piv-based_2013}, for which methodologies such as Particle Image Velocimetry (PIV) have become quite mature and able to supply both planar and volumetric velocity sets in either snapshots or time-resolved sequences.  

Examining the Navier-Stokes equations it is quickly apparent that, mathematically at least, the computation of pressure $P$ from a known velocity field $\vec u$ and density field $\rho$ should be a straightforward process. For this introduction we will concentrate on incompressible flows, but similar methods can be applied to compressible flows as well (see \citet{vanOudheusden2008} and the discussion in Section \ref{sec:DF_SPIV}).  Here, $\rho$ is density, $t$ is time, and $\mu$ is dynamic viscosity.
\begin{equation}
	\nabla P = -\rho \left( \frac{\partial \vec u}{\partial t} +  (\vec u \cdot \nabla) \vec u \right) + \mu \left(\nabla ^2 \vec u\right) = \vec f \left(\vec u \right)
\end{equation}

The most obvious approach would be to calculate the instantaneous pressure gradients from estimates of the material derivative of the velocity field, either including or ignoring the viscous terms. These estimates are lumped as a ``source term'' $\vec{f}(\vec{u})$, and will have a different form when the flow is compressible or when the average pressure is sought. Integrating these pressure gradients in space from some known starting pressure to the remaining points in the experimental domain would then allow for the estimation of the pressure field up to a constant.

Since the pressure field is a scalar field, the integration of the pressure gradient should be path-independent. Therefore, it is a sensible choice to build an algorithm that integrates the pressure gradient field from different directions and averages the results to minimize the effect of localized experimental error.
These methods evolved from averages of a small number of redundant paths toward efforts to average all possible straight line integrals within the domain, with the latter forms often being known as omnidirectional pressure integration schemes \citep{liu_error_2020}. Iterative application of this family of techniques allows convergence of the solution without requiring explicit specification of boundary conditions.

Alternately, if the divergence of the above equation is taken, a Poisson equation relating the pressure field to velocities can also be developed, which has the advantage of having many efficient computational implementations for its solution, assuming that the boundary conditions are known.  

\begin{equation} \label{eq:Poisson}
	\nabla^2	 P = \nabla \cdot \vec f \left(\vec u \right)
\end{equation}

In a series of papers, Pan, Whitehead and their collaborators demonstrated that, for Poisson solvers, experimental error on the final solution interacts with the geometry of the domain, how the measurements are sampled, the type of boundary conditions chosen, and the location and type of the error that is present in the data, among other factors \citep{pan_error_2016,faiella_error_2021,nie_error_2022}. Many of these same concerns apply to other solvers as well, as demonstrated in recent work by \citet{Sperotto2022} and \citet{Zhang2022}.
 
 Thus, in practice solving the Pressure-Poisson equation could lead to solutions that are sensitive to error in the measured velocity fields, with high levels of noise amplification.
\cite{charonko_assessment_2010} showed that in general, Poisson-based solvers are more sensitive to measurement error than omnidirectional integration methods.  
In particular, work by \cite{faiella_error_2021} showed that the sensitivity to experimental error in Poisson-based solvers can largely be attributed to the boundary-value problem nature of the approach, with the known pressure boundaries (Dirichlet) being superior to the more commonly-used pressure gradient boundaries (Neumann). 
Additional work by \cite{liu_error_2020} showed that omnidirectional schemes control this effect by iteratively converging the most-physically relevant pressure boundary conditions based on the entire measured domain before solving for the interior points in the last iteration. Specifically, when the boundary conditions extracted from the final iteration of a converged omnidirectional integration solver were used as Dirichlet boundaries in a Poisson solver, the results were essentially identical to the omnidirectional method. This suggests that it is the improved handling and minimization of the pressure error on the domain boundaries that is the reason that the omnidirectional integration schemes are in general more accurate than Poisson solvers.

However, modern fluid dynamics experiments are moving toward volumetric measurement techniques which better capture the full three-dimensional behavior of the flow and which, in theory, should also provide better estimates of the true pressure field.
Unfortunately, even in 2D the omnidirectional pressure integration schemes are considerably more expensive than Poisson solvers, and in 3D the difference in computational and memory requirements becomes even more pronounced, and often prohibitive.
As a result, few researchers have embraced omnidirectional pressure integration despite its potential superiority in accuracy.

In this paper, we will explore an alternate perspective on the mathematics behind the omnidirectional pressure integration schemes. We will then show how the approach can be recast as an implicit, Poisson-like problem while preserving the iterative convergence of the boundary conditions that powers its improved accuracy.
The approach will be demonstrated in synthetic and experimental fields of varying complexity, and is adaptable to both 2D and 3D flow data. 
Error analysis indicates that the new method demonstrates equivalent or better error levels as the conventional omnidirectional pressure integration, while maintaining computational efficiency equivalent, per iteration, to a Poisson solver. Finally, details will be provided on how to apply this technique to arbitrarily masked grids, where regions of the grid may contain missing data due to lack of particles, laser reflections or occlusions caused by the presence of an aerodynamic model.

\section{The Matrix-based Omnidirectional Integration Method} \label{MatrixOmni}

\subsection{Method Concept}  \label{sec:Concept}

In order to convert the explicit omnidirectional scheme into an implicit formulation, we first examine how the line integrals are constructed.  
For a discrete integration, the pressure at each new point, $P_C$, is computed from the pressure at the adjacent point, $P_a$, plus the pressure gradient along the path, $\vec r$.

\begin{equation} \label{eq:PressureIntegrator1}
	P_C = P_a + \int ^ {\vec x_C} _{\vec x_a} \frac{\partial P}{\partial \vec r} \cdot d\vec r
\end{equation}

\begin{figure*} [h]
	\centering
	\includegraphics[width=135mm]{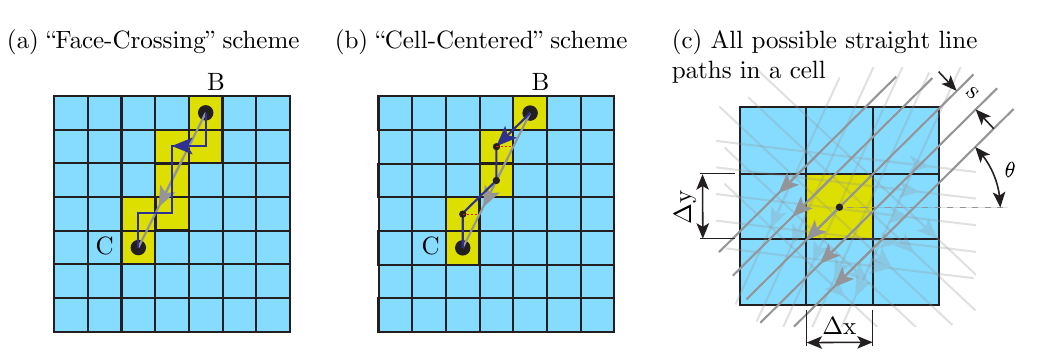}
	\caption{\centering Graphical description of the integration schemes discussed.}
	\label{fig:line-integration}
\end{figure*}

The collection of discrete cells with pressure gradient information (computed from velocity field data) that make up a path across the domain can be defined in multiple ways (Figure \ref{fig:line-integration}).
In this work, we will consider two approaches, shown here in 2D for simplicity. Both approaches considered assume a rectangular, evenly-spaced grid. Irregular grids will not be considered throughout this work. 

The first type of path, herein dubbed ``Face-Crossing scheme'' (Figure \ref{fig:line-integration} (a)), is made up of all cells that a given path $\vec r$ crosses, and all integrations are made across connected cell faces. The integral path starts from a boundary point $B$ and extends to one of the interior cells $C$.
The second type of path examined, dubbed ``Cell-Centered scheme'' (Figure \ref{fig:line-integration} (b)), consists of the cells with the closest centers.  
In this case, integrations may be across adjacent cell faces or diagonals. In 2D, if the ray angle $\theta$ is steeper than the cell diagonal (i.e., $\theta > \textrm{atan}(\Delta y/\Delta x)$), then the cell-centered integration path passes through each $y$ coordinate once and the $x$ coordinate is rounded to the nearest cell for a given ray. Conversely, if the ray angle is less steep than the diagonal ($\theta < \textrm{atan}(\Delta y/\Delta x)$), then the path passes through each $x$ coordinate once and the $y$ coordinate is rounded instead.

For the full omnidirectional pressure calculation, the pressure in any cell $C$ will be found from the average of all line integrals that cross the cell (Figure \ref{fig:line-integration} (c)), which we will determine using the parallel-ray omnidirectional approach \citep{liu_instantaneous_2016}. 
In that method, line integrals of pressure are found for closely spaced rays separated by a shift $s$ and oriented along multiple angles $\theta$.

\begin{figure*} [h]
	\centering
	\includegraphics[width=135mm]{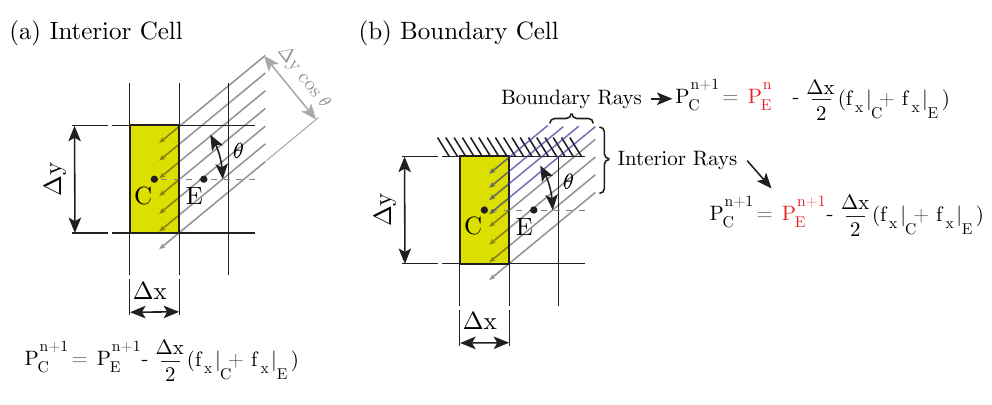}
	\caption{\centering (a) Definition of the ray counting process for the continuous integration case in an interior cell. (b) An adjacent cell can be treated as a boundary or as an interior point depending on whether the ray begins outside the boundary.}
	\label{fig:integration-details}
\end{figure*}

The proposed method extends the rotating parallel ray approach presented by \citet{liu_instantaneous_2016} by taking the limit where the number of line integrals is infinite. Thus, we consider all possible ray displacements $s$ and all possible ray angles $\theta$. As an example, let's analyze the specific case where the rays coming from the east cell $E$ are being counted, as depicted in Figure \ref{fig:integration-details}.

If the cell $E$ is an interior cell (i.e., none of the rays at any angle $\theta$ come from a boundary), then the rays can be ``counted'' through the integration of a connectivity variable $A_E$:

\begin{equation}
	A_E=\int_{-\pi}^{\pi}\int_{-\infty}^{\infty}R_{CE}(s,\theta) ds d\theta
\end{equation}

\noindent where $R_{CE}=1$ if $C$ and $E$ are connected by a path given some $s$ and $\theta$; and $0$ otherwise. The inner integral in $s$ can be replaced by the projection of the face connecting $C$ and $E$ (in 2D, the interface is an edge), which has height $\Delta y$ in the direction $\theta$. Also for the $E$ cell, the valid integration bounds for $\theta$ are $-\pi/2 < \theta <\pi/2$, and the integral $A_E$ becomes:

\begin{equation} \label{eq:AE_Interior}
	A_E^{n+1}=\int_{-\pi/2}^{\pi/2}\Delta y \cos \theta d\theta = 2\Delta y \; \; \textrm{(Interior Cell)}
\end{equation}

Note in Figure \ref{fig:integration-details} (a) that all rays coming from the cell $E$ implement the same equation: $P_C^{n+1}=P_E^{n+1} - (\Delta x/2) [f_x(C) + f_x(E)]$, where $f_x(j)$ is the $x$ component of the source function evaluated at the cell $j$ and the pressure $P_j^{n+1}$ represents the pressure of the cell $j$ evaluated at the $(n+1)^{th}$ iteration of the algorithm. The discretization of the source term $\vec{f}$ in this implementation is a second-order accurate approximation at the cell face, which is an averaging between the values at the center of the cells $C$ and $E$. For an interior cell, Equation \ref{eq:AE_Interior} defines the ray count $A_E^{n+1}$, whereas $A_E^{n}=0$ because no pressure values from the previous iteration are required.

However, when the cell $E$ shares one or more faces with a boundary, then the rays intersecting the boundary faces will effectively begin at the cell $E$, meaning cell $E$ is the boundary cell where the integration starts. Therefore, the pressure value used for cell $E$ has to come from the previous ($n^{th}$) iteration, as depicted in Figure \ref{fig:integration-details} (b). For the ``zeroth'' iteration, the values can be initialized to zero \citep{liu_error_2020} or any other arbitrary value. Since for the boundary cell case some rays use pressure values from the $(n+1)^{th}$ iteration and some rays use values from the $n^{th}$ iteration, the ray count $A_E^i$ (where $i$ is the iteration number) will be different than Equation \ref{eq:AE_Interior}. Depending on the location of cell $C$, the values of $A_E^i$ will vary. As an example, the values for the boundary topology depicted in Figure \ref{fig:integration-details} (b) are:

\begin{equation} \label{eq:AE_Boundary_example}
	\begin{split}
	A_E^{n}=\int_{0}^{\textrm{\scriptsize{atan}}(\Delta y/\Delta x)}\Delta x \sin \theta d\theta + \int_{\textrm{\scriptsize{atan}}(\Delta y/\Delta x)}^{\pi/2} \Delta y \cos \theta d\theta \\
	A_E^{n}= \Delta x + \Delta y - \sqrt{\Delta x^2 + \Delta y^2}
	\end{split}
\end{equation}

\begin{equation} \label{eq:AE_Boundary_example2}
	A_E^{n+1}=2\Delta y - A_E^{n} = \Delta y - \Delta x + \sqrt{\Delta x^2 + \Delta y^2}
\end{equation}

Given the forms of the integrals, the ``ray count'' $A_j^i$ is effectively the average of edge length ($\Delta x$ or $\Delta y$) projected at all angles $\theta$ that come from cell $j$ and use the pressure value from iteration $i$. For the ``face-crossing'' scheme, we can repeat the application of Equation \ref{eq:AE_Interior} to the other faces to find the total projection area from all rays at all angles to be:

\begin{equation} \label{eq:Atot2D}
	A_{tot}=\sum_j \sum_i A_j^i = 4(\Delta x + \Delta y)
\end{equation}

\noindent since according to Equation \ref{eq:AE_Interior}, $(A_E^{n}+A_E^{n+1})=(A_W^{n}+A_W^{n+1})=2\Delta y$ and similarly $(A_N^{n}+A_N^{n+1})=(A_S^{n}+A_S^{n+1})=2\Delta x$ due to the symmetry of the problem.

We can then define an implicit equation that implements the averaging process from all rays by performing a weighted average of the finite-difference approximations of Equation \ref{eq:PressureIntegrator1} from all adjacent cells:

\begin{equation} \label{eq:ExplicitEqn_FaceCrossing}
	P_C^{n+1}= \sum_i \sum_j w_j^i P_j^i - \sum_{j} \Delta j (w_{j}^{n+1} + w_{j}^{n}) \frac{f_j(j)+f_j(C)}{2}
\end{equation}

\noindent where $P_C^{n+1}$ is the updated value of the pressure at a given cell $C$ at the next $(n+1)^{th}$ iteration, and the iteration $i$ can take the values $i=\{n,n+1\}$. The cell $j$ can take the values $j=\{C,E,W,N,S,F,B\}$ for the ``face-crossing'' scheme, representing the cardinal directions East, West, North, South, Front and Back, respectively. For the cell-centered scheme, the cardinal directions can also take diagonal values $j=[\{p\},\{pq\},\{pqr\}]$ representing all 27 cells in a 3$\times$3 cube surrounding cell $C$, including itself. $j$ is also a directional subscript, so $\Delta E=\Delta x$ and $\Delta W=-\Delta x$, and similarly for the other orthogonal and diagonal directions. The source function $f_j$ (at direction $j$) in the last summation is averaged between cells $j$ and $C$ to approximate the source value at the corresponding face, following a second-order accurate approximation.

Equation \ref{eq:ExplicitEqn_FaceCrossing} simply describes the ray averaging process in the omnidirectional integration approach. The weights $w_j^i$ represent the fraction of the rays that implement each sub-equation (Equation \ref{eq:PressureIntegrator1}), utilizing the values of pressure from the appropriate neighboring cells depending on which direction the ray is coming from. These weights are defined as:

\begin{equation}  \label{eq:weights}
	w_j^i=\frac{A_j^i}{A_{tot}}
\end{equation}

\noindent except for $w_C^{n+1}$, which is defined as:

\begin{equation} \label{eq:wcnp1}
	w_C^{n+1}=0
\end{equation}

This is because $P_C^{n+1}$ is already counted in the left-hand side of Equation \ref{eq:ExplicitEqn_FaceCrossing}. The remaining weights are computed from the ray integrals $A_j^i$ as previously discussed.

Note that Equation \ref{eq:ExplicitEqn_FaceCrossing} can be defined for all cells in the domain, forming a sparse matrix form:

\begin{equation}  \label{eq:MatrixEquation}
	[W^{n+1}] \{P^{n+1}\} =	[W^{n}] \{P^{n}\} +\{S\}
\end{equation}

\noindent where $[W^{i}]$ are weight matrices and $\{P^{i}\}$ are the pressure vectors at iteration $i$. The source term vector $\{S\}$ represents the last summation in Equation \ref{eq:ExplicitEqn_FaceCrossing} and is computed only once based on the velocity measurements from PIV. For a minimal example of how Equation \ref{eq:MatrixEquation} would be constructed for a 3$\times$3 domain, the reader is directed to the appendix of this manuscript. Furthermore, the weight matrices $[W^{i}]$ can be precomputed for a given grid topology and can include the effect of cells with missing data (say, from regions not illuminated by the laser or containing occlusions/reflections). A method to consider any topology  of missing cells will be detailed in the following sections.

\begin{algorithm}
	\caption{Outline of the matrix-based omnidirectional integration algorithm}\label{alg:omnidirectional}
	\begin{algorithmic}
	\REQUIRE Source field $\vec{f}$ from PIV data \\
	\textbf{Compute:} \textit{Source vector} $\{S\}$ (Eq. \ref{eq:ExplicitEqn_FaceCrossing}, terms using $\vec{f}$) \\
	\textbf{Compute:} $c_{jk}$ and $w_i^j$ (details in Section \ref{sec:WeightComponents}) \\
	\textbf{Compute:} \textit{Weight matrices} $[W^{i}]$ (Eq. \ref{eq:weights}) \\
	\textbf{Initialize:} $\{P^{n=0}\}=0$; \\
		
	\WHILE{\textit{Residual} $>$ \textit{Threshold}} 
	\STATE{\textbf{Solve:} Eq. \ref{eq:MatrixEquation} for $\{P^{n+1}\}$} 
	\STATE{\textbf{Compute:} \textit{Residual} (Eq. \ref{eq:ResidualPressure} or \ref{eq:ResidualMomentum})}
	\STATE{$\{P^{n}\}\gets \{P^{n+1}\}$}
	\ENDWHILE
	
	\end{algorithmic}
\end{algorithm}

The solution method proposed follows the outline provided in Algorithm \ref{alg:omnidirectional}. This algorithm was implemented in Matlab as a CPU code (only 2D) and in CUDA-C++ as a GPU code (2D and 3D). For the Matlab implementation, the weight matrices $[W^{i}]$ are stored as sparse matrices and  Equation \ref{eq:MatrixEquation} is solved using Matlab's default linear equation system solver (\textit{mldivide} or \textit{``backslash'' operator}). In the GPU CUDA-C++ implementation the Conjugate Gradient (CG) method is used to solve Equation \ref{eq:MatrixEquation}. 

For the GPU implementation, a total of 16 scalar fields with the mesh size are required (7 for the CG solver, 6 for the weight matrices, 3 for the source field $x,y,z$ components). Stored as doubles, the code requires $16\cdot 8N$ bytes of GPU memory, where $N=N_x N_y N_z$ is the total number of cells in the domain. On the Nvidia RTX A4000 GPU used in this work, which has 16GB of memory and retails at $\sim$\$1000 at the time of writing, a mesh size of at most $N=512^3 = 134$ million cells can be computed. Since the CG solver could potentially compute the weight matrices on-the-fly, it is possible to trade off computational time to increase the size of the domain, requiring only 10 scalar fields to be stored in memory. This possibility was not explored, since a domain of 134 million cells was deemed sufficient for most Tomographic PIV applications currently being considered by the experimental fluids community.

\subsection{Weight Computation for Arbitrary Boundary Shapes} \label{sec:WeightComponents}

\begin{figure*} [h]
	\centering
	\includegraphics[width=95mm]{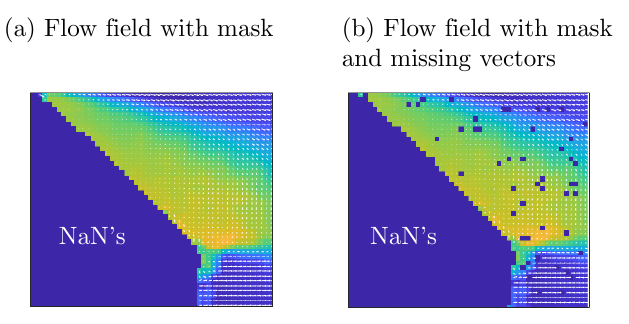}
	\caption{\centering Simplified example of a flow field with masking (a) and masking plus missing vectors (b) [Not considered in this work]}
	\label{fig:mask}
\end{figure*}

When working with experimental data from PIV/PTV, it is common to have cells in the 2D or 3D grid that are masked out due to missing data. Missing PIV vectors can occur when laser reflections from a model surface are present in the field of view; when a model surface is occluding the view of the flow; or when the laser sheet is not sufficiently expanded to illuminate the entire field of view of the cameras. Alternatively, in Stereoscopic PIV and Tomographic PIV there can also be missing vectors where the views from all cameras do not overlap. A typical flow field with vectors masked out due to the presence of an aerodynamic model is exemplified in Figure \ref{fig:mask} (a). 

Vectors could also be missing inside the field of view due to post-processing or thresholding (say, if the correlation coefficient does not meet a validation criterion), as shown in Figure \ref{fig:mask} (b). This case will \textbf{not} be considered in this work, vectors missing due to vector post-processing will be filled with an interpolation algorithm. However, the approach that will be described in this section is agnostic to the nature of the missing vectors and should still handle these cases. Note Figure \ref{fig:mask} encodes the missing vectors as NaN (Not a Number) values, shown as a deep blue color in the contours. NaN values will encode missing vectors throughout this manuscript.

When vectors are missing due to masking (Figure \ref{fig:mask} (a)), the integration of a pressure ray needs to be performed starting at the edges of the masked out region. Furthermore, the masked region can have concave regions where some of the rays will start and end without ever reaching the borders of the domain. Therefore, a more general approach to compute the weights $w_i^j$ considering arbitrarily masked border shapes is necessary to deal with realistic PIV datasets, especially in 3D volumes where the masked borders can have a large number of combinations of missing data topologies that would be intractable to handle manually in code.

Here a method will be introduced for the computation of the weights for arbitrary domain shapes. Only the ``face-crossing'' scheme will be described in the body of this manuscript, for both 2D and 3D grids. The ``cell-centered'' scheme is described in the supplementary material only for 2D grids, as the 3D grid integrals were exceedingly cumbersome to compute in the ``cell-centered'' scheme with arbitrary boundaries.

\subsubsection{Arbitrary boundary weights for ``face-crossing'' scheme in 2D}

\hfill \newline

\begin{figure*} [h]
	\centering
	\includegraphics[width=145mm]{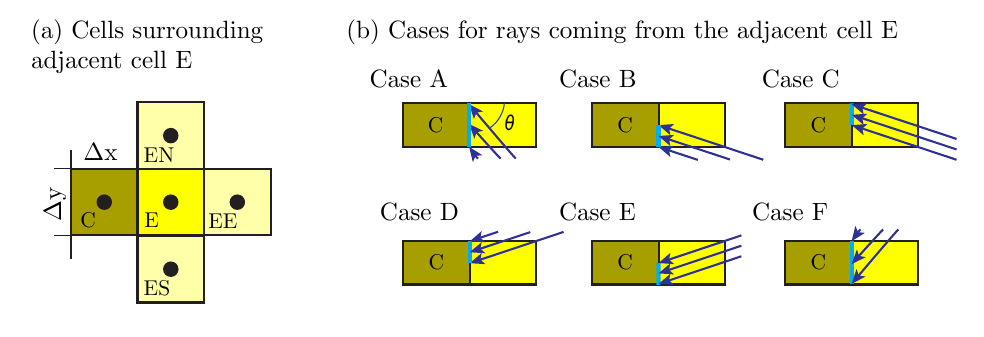}
	\caption{\centering Graphical description of the arbitrary weight computation considering the cells neighboring the adjacent cell $E$.}
	\label{fig:2dfacecrossing}
\end{figure*}

In order to compute the weights corresponding to interior cells with potentially arbitrary boundaries, let's consider the source field $\vec{f}$ has already been computed and it contains NaN values at some cells. If a cell has a NaN neighbor, rays coming from the NaN direction must be treated as boundaries and therefore must be assigned to $w_j^n$ (i.e., the $n^{th}$ iteration). In the face crossing scheme, a ray coming from the east ($E$) cell will have a path that passes through the east cell as well as one and only one of its neighbors (i.e., it must pass through either $EN$, $EE$ or $ES$ in Figure \ref{fig:2dfacecrossing} (a)).

We can define a boolean value $b_j$ for each cell, defining whether there is missing data in the corresponding cell for the source field $\vec{f}$:

\begin{equation}
	b_j=\begin{cases}
		0 \;\; \textrm{if } \vec{f}(j) \neq \textrm{NaN} \\
		1 \;\; \textrm{if } \vec{f}(j) = \textrm{NaN}
	\end{cases}
\end{equation}

We also define $\tilde{b}_j=1-b_j$ as the boolean NOT operation on $b_j$ to define where data is available instead.

Thus, when considering all possible ray angles $\theta$, we need to assess six possible cases depending on where the rays come from and whether the corresponding adjacent cells have missing data or not. The cases are enumerated in Figure \ref{fig:2dfacecrossing} (b). For example, case A represents rays coming at an angle such that all rays come from the $ES$ cell but some rays miss the $C$ cell (and therefore must not be counted). Case B represents rays coming at an angle such that all rays coming from the $ES$ cell also enter the $C$ cell. Case C is the complementary case, where the rays at the same angle come from the $EE$ cell and cross the $C$ cell. Cases D, E and F are complementary cases to C, B and A, respectively, but involving the $EN$ cell instead.

For each of the cases, a different mathematical expression representing the projection of the edge highlighted in blue can be defined. The total ray projection for the $(n+1)^{th}$ iteration then becomes:

\begin{equation} \label{eq:GiantEquation2D}
	\begin{split}A_E^{n+1}= \tilde{b}_E \bigg[ \underbrace{\tilde{b}_{ES} \int_{-\pi/2}^\alpha \Delta y \cos\theta d\theta}_{\textrm{Case A}} +  \underbrace{\tilde{b}_{ES} \int_{-\alpha}^0 \Delta x \sin|\theta| d\theta}_{\textrm{Case B}} +  \underbrace{\tilde{b}_{EE} \int_{-\alpha}^0 (\Delta y \cos\theta - \Delta x \sin |\theta|) d\theta}_{\textrm{Case C}}  \dots  \\
	+\underbrace{\tilde{b}_{EE} \int_{0}^\alpha (\Delta y \cos\theta - \Delta x \sin \theta) d\theta}_{\textrm{Case D}}  +\underbrace{\tilde{b}_{EN} \int_{0}^\alpha \Delta x \sin|\theta| d\theta}_{\textrm{Case E}} +\underbrace{\tilde{b}_{EN} \int_{\alpha}^{\pi/2} \Delta y \cos\theta d\theta}_{\textrm{Case F}}	\bigg]
	\end{split}
\end{equation}

\noindent where $\alpha = \textrm{atan}(\Delta y/\Delta x)$. Note that $\tilde{b}_E$ multiplies the entire expression, to indicate that if there is missing data in the $E$ cell then this entire weight evaluates to zero. When the data in the $E$ cell is missing, then the current cell being evaluated (cell $C$) must be the starting point for all rays going westward and therefore is itself a boundary. Therefore, there must also be a boundary contribution from the cell $C$ of the form:

\begin{equation} \label{eq:Acn}
	A_C^n = 2\Delta y(b_E + b_W) + 2\Delta x(b_N+b_S)
\end{equation}

This form considers all four cells surrounding cell $C$, and the constants $2\Delta x$ and $2\Delta y$ are calculated the same way as shown in Equation \ref{eq:AE_Interior}. We can evaluate the integrals in Equation \ref{eq:GiantEquation2D} to obtain the expressions:

\begin{equation} \label{eq:AEnp1}
	A_E^{n+1}= \tilde{b}_E [c_{xy} (\tilde{b}_{ES} + \tilde{b}_{EN}) + c_{xx} \tilde{b}_{EE}]
\end{equation}

\noindent where the constants $c_{jk}$ are defined as:

\begin{equation} \label{eq:cxx2D}
	\begin{split}c_{xx}=2(\Delta d - \Delta x) \\
		c_{yy}=2(\Delta d - \Delta y) \\
		 c_{xy}=\Delta x + \Delta y - \Delta d
	\end{split}
\end{equation}

\noindent and $\Delta d = \sqrt{\Delta x^2 +  \Delta y^2}$. These constants come from evaluating the integrals in Equation \ref{eq:GiantEquation2D}.

Similarly, the ray integral corresponding to the boundary values for iteration $n$ has the following form:

\begin{equation} \label{eq:AEn}
	A_E^{n}= \tilde{b}_E [c_{xy} (b_{ES} + b_{EN}) + c_{xx} b_{EE}]
\end{equation}

To compute the weights corresponding to the west ($W$) cell, one simply swaps $E$ and $W$ in Equations \ref{eq:AEnp1} and \ref{eq:AEn}. For the north ($N$) and south ($S$) cells, the constant $c_{xx}$ also needs to be swapped to $c_{yy}$:

\begin{equation} \label{eq:ANnp1}
	A_N^{n+1}= \tilde{b}_N [c_{xy} (\tilde{b}_{EN} + \tilde{b}_{WN}) + c_{yy} \tilde{b}_{NN}]
\end{equation}

\begin{equation}  \label{eq:ANn}
	A_N^{n}= \tilde{b}_N [c_{xy} (b_{EN} + b_{WN}) + c_{yy} b_{NN}]
\end{equation}

For a uniform rectangular grid, the constants $c_{xx}$, $c_{yy}$ and $c_{xy}$ can be trivially precomputed using Equation \ref{eq:cxx2D}. Then, the ray counting process to determine $A_j^i$ reduces to a conditional sum for each cell in the domain using the values $b_j$ that define whether data for the function $\vec{f}$ is available in the corresponding neighboring cells. The weights $w_j^i$ then are found using Equation \ref{eq:weights}.

\subsubsection{Arbitrary Boundary Weights for ``Face-Crossing'' Scheme in 3D} \label{sec:FaceCrossing3D}

\hfill \newline

In the three-dimensional case, the computation of weights is significantly more involved, and therefore full detail of the integrals computated will be provided in the supplementary material. In this manuscript, we describe the key steps involved in the weight computation in 3D. As the rays now come from all spherical angles, we have to compute double integrals with curved bounds. The complexity of the integrals is also increased because the projections of the neighboring faces onto the faces of the cell $C$ are no longer lines, but triangles, trapezoids or rectangles depending on the values of the spherical angles.

The problem is still tractable, as will be demonstrated herein. Not all integrals will have a closed form, thus requiring them to be numerically computed. However, this computation only occurs once and only a handful of coefficients needs to be stored in memory. Similarly to the 2D case, this method has the advantage of treating all possible combinations of boundary conditions in 3D, which is combinatorially intractable. In the 2D ``face-crossing'' scheme, there are $2^{12} = 4096$ possible combinations of boundary conditions. In 3D, the number of combinations is $2^{32}\approx 4$ billion. Evidently, most combinations in 3D would be degenerate cases, but it is still better to have a system that can treat any possible combination of boundary conditions than having to treat each boundary condition possibility within a pre-programmed case structure.

\begin{figure*} [h]
	\centering
	\includegraphics[width=130mm]{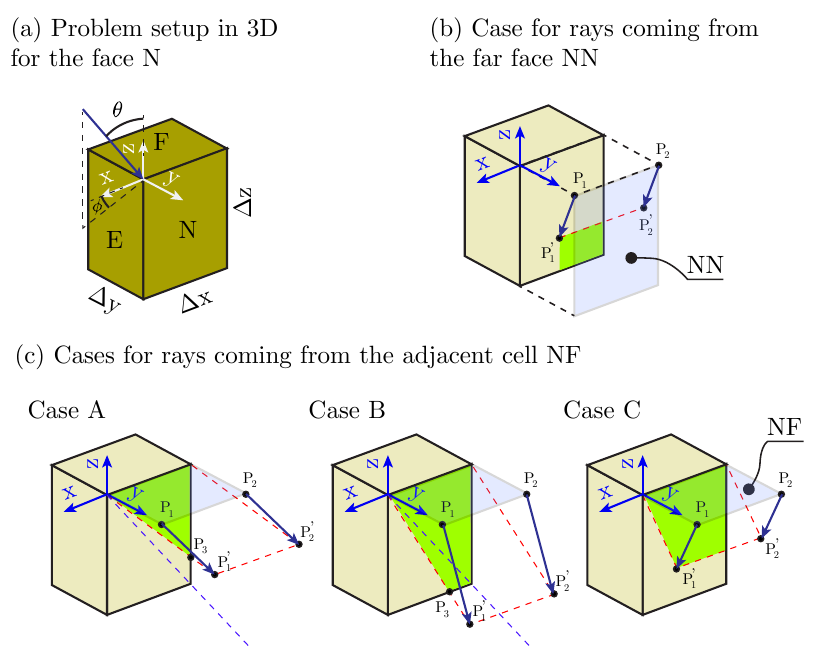}
	\caption{\centering Coordinate system definition (a), projection of far face $NN$ on face $N$ (b) and projection of dihedral face $NF$ on face $N$ (b).}
	\label{fig:3dfacecrossing}
\end{figure*}

First, we will find the value of $A_{tot}$, considering all rays coming from all directions. For this, we will define a spherical coordinate system with polar angle $\theta$ and azimuthal angle $\phi$, according to the right-handed coordinate system described in the ISO 80000-2:2019 convention. The definitions of the angles $\theta$ and $\phi$ are depicted in Figure \ref{fig:3dfacecrossing} (a).

Counting the total number of rays crossing the north face ($N$), which has a positive $y$ normal, is straightforward. Defining a unit vector in the ray direction $\hat{r}$:

\begin{equation}
	\hat{r} = 
	\begin{bmatrix}
		\sin{\theta} \cos{\phi} \\ \sin{\theta} \sin{\phi} \\ \cos{\theta}
	\end{bmatrix}
\end{equation}

The projection of the face $N$ in the direction $\hat{r}$ is found by multiplying the surface area of $N$ ($\Delta x \Delta z$) by the dot product $\hat{y}\cdot\hat{r}$. To find the total projection from all spherical angles $A_N$, one needs to integrate:

\begin{equation}
	A_N = 4 \int_0^{\pi/2} \int_0^{\pi/2} \Delta x \Delta z \big(\hat{y} \cdot \hat{r}\big) \sin{\theta} d\theta d\phi
\end{equation}

The factor of $4$ is present due to the symmetry of the problem and the defined bounds of the angles $\theta$ and $\phi$. Alternatively, one could make the upper bounds of both angles $\pi$ and remove the factor of $4$, yielding the same results.This integral can be evaluated analytically:

\begin{equation}
	A_N = 4 \Delta x \Delta z \int_0^{\pi/2} \int_0^{\pi/2}  \big(\sin{\theta} \sin{\phi} \big) \sin{\theta} d\theta d\phi = \pi \Delta x \Delta z 
\end{equation}

Given the symmetry of the problem, the total projection for the other directions ($A_E, A_W, A_F, A_B, A_S$) can be found by swapping the grid spacings $\Delta x$ and $\Delta z$ by the appropriate values. Doing this yields a total projection area of:
\begin{equation}
	A_{tot} = 2\pi (\Delta x \Delta y +\Delta x \Delta z +\Delta y \Delta z )
\end{equation}

Within the same theme of utilizing a boolean value $b_j$ to define whether a cell $j$ has NaN as a source term $\vec{f}$ (and therefore must be treated as a boundary), we can evaluate the area integrals for the interior and boundary contributions in 3D in a similar form as it was performed in 2D:

\begin{equation}
	\begin{split}
	A_N^{n+1}= \tilde{b}_N [c_{xy} (\tilde{b}_{EN} + \tilde{b}_{WN}) + c_{yz} (\tilde{b}_{NF} + \tilde{b}_{NB}) + c_{yy} \tilde{b}_{NN}] \\	
	A_N^{n}= \tilde{b}_N [c_{xy} (b_{EN} + b_{WN}) + c_{yz} (b_{NF} + b_{NB}) + c_{yy} b_{NN}]
	\end{split}
\end{equation}

The pattern is also the same, where ($N$ and $S$), ($W$ and $E$) and ($F$ and $B$) are swappable in the expressions above, and the corresponding constants $c_{jk}$ must correspond to the appropriate directions (i.e., $y$ for $N$ and $S$,  $x$ for $E$ and $W$, and  $z$ for $F$ and $B$). This means a total of 6 constants $c_{jk}$ must be calculated. 

For the far faces ($j=k$), the area projection is displayed in Figure \ref{fig:3dfacecrossing} (b), where $P_1$, $P_2$ are points on one of the edges of the far face and $P_1^\prime$, $P_2^\prime$ are the projections of these points on the $N$ face through the $\hat{r}$ direction. Generalizing the projection for arbitrary coordinates $(j, k, l)$ results in the following integral that finds the contribution of the far face to the area projection $A$:

\begin{equation} \label{eq:FarFaceEquation}
	\begin{split}
	c_{jk} (j=k) = \int_{\textrm{atan}(\Delta j/\Delta \ell_2)}^{\pi/2} \pi(\Delta \ell_2 \Delta \ell_{1,s} - \Delta j \Delta \ell_{1,c}) + \dots \\	
	2\bigg[\Delta j\Delta \ell_{1,c} \;\textrm{atan}\frac{\Delta j}{\Delta \ell_{1,s}} - \Delta \ell_2\Delta \ell_{1,s} \;\textrm{atan}\frac{\Delta j}{\Delta \ell_{1,s}} \bigg] d\phi
\end{split}
\end{equation}

\noindent where $\Delta j$ is the grid spacing in the $j$ direction and $\Delta \ell_1$ and $\Delta \ell_2$ are the grid spacings in the remaining two directions.  Furthermore, $\Delta \ell_{1,c} = \Delta \ell_1 \cos{\phi}$ and $\Delta \ell_{1,s} = \Delta \ell_1 \sin{\phi}$. Although $\Delta \ell_1$ and $\Delta \ell_2$ do not seem to be interchangeable in Equation \ref{eq:FarFaceEquation}, the result from the integral is identical when $\Delta \ell_1$ and $\Delta \ell_2$ are swapped. Further details on how to achieve the integral form of Equation \ref{eq:FarFaceEquation} can be found in the supplementary material.

For the faces forming dihedrals (i.e., for $c_{jk}$ where $j\neq k$), the projections on the $N$ face can form triangles or trapezoids, depending on the values of the angles $\theta$ and $\phi$. There are three cases that must be considered, enumerated in Figure \ref{fig:3dfacecrossing} (c). These cases result in four integrals:

\begin{equation}
	c_{jk} (j\neq k) = I_{A} + I_{B_1}+ I_{B_2}+ I_{C}
\end{equation}

\begin{equation} \label{Equation_IAc}
	I_{A} = \frac{\Delta \ell^2}{2} \textrm{atanh} \bigg( \frac{\Delta j^2 \Delta k^2}{2\Delta \ell^4 + 2\Delta \ell^2 \Delta j^2 +2\Delta \ell^2 \Delta k^2 + \Delta j^2 \Delta k^2}\bigg)
\end{equation}

\begin{equation}\label{Equation_IB1c}
	\begin{split}
			I_{B_1} =  \int_0^{\textrm{atan}(\Delta j/\Delta \ell)} \bigg\{\Delta \ell \Delta k_s\bigg( \textrm{atan} \frac{\Delta \ell}{\Delta k_c} - \frac{\Delta \ell \Delta k_c}{\Delta \ell^2 + \Delta k_c^2} \bigg) + \dots \\ 
			\Delta k_c \Delta k_s\bigg[\ln \bigg(\frac{\Delta k_c}{\sqrt{\Delta k_c^2 + \Delta \ell^2}}\bigg) + \frac{\Delta \ell^2}{2(\Delta \ell^2 + \Delta k_c^2) }\bigg]\bigg\} d\phi
	\end{split}
\end{equation}

\begin{equation}\label{Equation_IB2c}
	\begin{split}
			I_{B_2} =  \int_{\textrm{atan}(\Delta j/\Delta \ell)}^{\pi/2} \bigg\{ \Delta \ell \Delta k_s\bigg( \textrm{atan} \frac{\Delta j}{\Delta k_s} - \frac{\Delta j \Delta k_s}{\Delta j^2 + \Delta k_s^2} \bigg) + \dots \\ 
			\Delta k_c \Delta k_s\bigg[\ln \bigg(\frac{\Delta k_s}{\sqrt{\Delta k_s^2 + \Delta j^2}}\bigg)  + \frac{\Delta j^2}{2(\Delta j^2 + \Delta k_s^2)}\bigg]\bigg\} d\phi
	\end{split}
\end{equation}

\begin{equation} \label{Equation_ICc}
	I_{C}=\frac{1}{2} \int_{\textrm{atan}(\Delta j/\Delta \ell)}^{\pi/2} \frac{\Delta j \Delta k^2 \sin{\phi} \big(2\Delta \ell \sin{\phi} - \Delta j \cos{\phi}\big)}{\Delta j^2 + \Delta k^2 \sin^2 \phi} d\phi
\end{equation}

\noindent where $\Delta \ell$ represents the grid spacing in the direction not considered ($\ell \neq \{j,k\}$) and $\Delta k_c = \Delta k \cos{\phi}$ and $\Delta k_s = \Delta k \sin{\phi}$.

Note Equation \ref{Equation_IAc} has a closed form, whereas the remaining integrals need to be computed numerically. The forms provided are the simplest analytical expressions the authors were able to derive. Although not straightforwardly evident at a first glance, $c_{jk}=c_{kj}$. This is because the projection of face $j$ onto face $k$ for any viewing angle always has the same area as the projection of face $k$ onto face $j$. Once again, further details on the computation of the integrals can be found in the supplementary material. 

Although the integrals in Equations \ref{Equation_IB1c}, \ref{Equation_IB2c}, \ref{Equation_ICc} and \ref{eq:FarFaceEquation} must be numerically computed, this computation only needs to be performed once prior to running the matrix omnidirectional solver. Computing all integrals to $1\times10^{-7}$ precision takes about 3 seconds for arbitrary $\Delta x$, $\Delta y$ and $\Delta z$, which was deemed a reasonable investment for solving an arbitrarily-spaced fully three-dimensional grid. For common ratios of $\Delta x$, $\Delta y$ and $\Delta z$, these integrals can also be precomputed and stored. A special case is shown in the next section.

\subsubsection{3D ``Face-Crossing'' Constants for Isotropic Grids}

\hfill \newline

For the special case where the grid spacing is equal in all directions (i.e., $\Delta x = \Delta y = \Delta z = \Delta$), we can perform further simplification of the integrals and attain a closed form for $c_{jk}$, which then can be added to the solver code as a fixed constant for isotropic grids. Below we can find the values of these constants:

\begin{equation} \label{ConstantFarFace}
	\begin{split}
	 c_{jk} (j=k) = \Delta^2 \bigg\{ \ln \bigg(\frac{4}{3} \bigg) - \sqrt{2} \; \textrm{atan} (\sqrt{2}) +\frac{\sqrt{2}}{2} \textrm{atan} \bigg(\frac{10\sqrt{2}}{23} \bigg) + \pi (\sqrt{2} -1) \bigg\} \\  \textrm{(Isotropic Grid)}
 \end{split}
\end{equation}

\begin{equation} \label{ConstantNearFaces}
	\begin{split}
	c_{jk} (j\neq k) = \Delta ^2 \bigg\{\frac{\pi}{2} - \frac{1}{4}\ln \left(\frac{4}{3}\right)-\frac{\sqrt{2}}{4} \textrm{atan}\bigg(\frac{10 \sqrt{2}}{23}\bigg)- \frac{\sqrt{2}}{2} \; \textrm{atan}(\sqrt{2})\bigg\}  \\  \textrm{(Isotropic Grid)}
	\end{split}
\end{equation}

Numerically, these constants have the following values:

\begin{equation} 
	c_{jk} (j=k) = 0.6277684243304 \Delta ^2\; \; \textrm{(Isotropic Grid)}
\end{equation}

\begin{equation} 
	c_{jk} (j\neq k) =0.6284560573148  \Delta ^2\; \; \textrm{(Isotropic Grid)}
\end{equation}

\section{Method Accuracy and Computational Complexity} 

In this section, the matrix-based omnidirectional pressure integration method described in Section \ref{MatrixOmni} will be tested with synthetic datasets where the ground-truth pressure field is known. Noise will be added to the velocity vectors to assess the uncertainty behavior of the pressure field solutions for the method proposed and for previous methods present in the literature. 

\subsection{Taylor Vortex (2D)} \label{sec:TaylorVortex}

First, a decaying Taylor vortex is considered, which was previously used by \citet{charonko_assessment_2010} to assess error propagation in pressure integration algorithms. The Taylor vortex has an analytical solution for pressure that enables the definition of a ground truth. The velocity field of the decaying Taylor vortex centered at the origin is defined in cylindrical coordinates as:

\begin{equation}
	u_\theta (r) = \frac{H}{8\pi} \frac{r}{\nu t^2} \exp \bigg(-\frac{r^2}{4\nu t}\bigg)
\end{equation}

\noindent where $u_\theta$ is the tangential velocity, $r$ is the radial position and $H$ is a measure of the angular momentum in the vortex. The pressure field for this vortex can be found by solving the Navier-Stokes equations:

\begin{equation}
	P (r) = -\frac{\rho H^2}{64\pi^2 \nu t^3}  \exp \bigg(-\frac{r^2}{2\nu t}\bigg) + P_\infty
\end{equation}

\noindent where $P_\infty$ is an integration constant and is defined as the far-field pressure. 

For this study, the parameters $H=10^{-6}$ $\textrm{m}^2$, $\nu=10^{-6}$ $\textrm{m}^2/\textrm{s}$, $\rho=1,000$ $\textrm{kg}/\textrm{m}^3$ were used. The characteristic scales for length ($L_0=\sqrt{H}$), velocity ($U_0=\nu/\sqrt{H}$), time ($T_0=H/\nu$) and pressure ($P_0=\rho U_0^2$) are also defined to normalize the values prior to sending the data to the solvers.

In order to generate velocity fields $\vec{u}_{noisy}$ contaminated with spatially uncorrelated noise, a noisy vector field $\vec{n}$ with unit magnitude and random angle is sampled from an uniform angular distribution and scaled by a factor $s$. It is then multiplied pointwise by the magnitude of the instantaneous velocity vectors $||\vec{u}||$, following a previous work \citep{charonko_assessment_2010}:

\begin{equation} \label{eq:NoisyVectorsTheta}
	\vec{u}_{noisy} = \vec{u} + s\vec{n} ||\vec{u}||
\end{equation}

The scaling factor $s$ is then varied between 0 and 0.1 (0\% and 10\% noise intensity) for this analysis. In all cases examined, a POD-based denoising scheme retaining only the most energetic modes in the time series was used to reduce noise in the velocity fields, following the previous work by \citet{charonko_assessment_2010}. Modes with less than 1\% of the total energy were discarded, resulting in approximately 6 modes retained for pressure reconstruction.

Five pressure reconstruction algorithms were tested in this part of the work, namely: 
\begin{enumerate}
	\item Pressure-Poisson solver with Neumann boundaries; \label{ppe}
	\item Omnidirectional integration, ``virtual boundary'' \citep{Liu2006}; \label{omni1liu}
	\item Omnidirectional integration, ``rotating parallel ray'' \citep{liu_instantaneous_2016}; \label{omni2}
	\item Matrix-based omnidirectional solver, ``face-crossing'' scheme; \label{matrixfc}
	\item Matrix-based omnidirectional solver, ``cell-centered' scheme; \label{matrixcc}
\end{enumerate}

For the Pressure-Poisson case (\ref{ppe}), we defined an additional Dirichlet boundary condition at one single mesh point at a corner of the domain to ensure the problem is well-defined. To define the source terms, the "standard" approach as defined in \citet{charonko_assessment_2010} was used. This method used the full Navier-Stokes equations for Neumann boundaries, and direct calculation of the divergence of each term for the source term, preserving the time-derivatives and viscous terms with no application of an incompressible flow assumption. This was chosen for consistency with the forms used in the omni-directional solvers, but as shown in \citet{charonko_assessment_2010} and \citet{nie_error_2022}, alternate forms that reduce numerical error propagation can sometimes result in better performance.

For the ``virtual boundary'' case (\ref{omni1liu}), a total of $2 N_x N_y$ rays were used in the boundary, whereas for the ``rotating parallel ray'' approach (\ref{omni2}) a ray spacing $\Delta s/\Delta x=0.4$ was used and $\Delta \theta$ was four times the value provided in formula (73) in \citet{liu_error_2020}.

To compare computational efficiency and accuracy of the algorithms tested, a 2D version of the algorithms described is implemented on a square domain of variable number of grid points containing the Taylor vortex flow field. The omnidirectional integration algorithms (\ref{omni1liu}) and (\ref{omni2}) are implemented in Matlab using the parallel computing toolbox, and 8 nodes of an Intel Xeon Gold 5218R CPU (2.10 GHz) were used to accelerate the computations. The algorithms requiring a matrix solver (\ref{ppe}), (\ref{matrixfc}), (\ref{matrixcc}) employed Matlab's default linear equation system solver (\textit{mldivide} or \textit{``backslash'' operator}). Finally, in the error tests considered in this section, a total of 26 Taylor vortex snapshots at times spanning $0.05<t/T_0<0.3$ using a $101^2$ grid that spans from $-3<x/L_0<3$ and $-3<y/L_0<3$ were employed.

\begin{figure*} [h]
	\centering \hspace*{-5mm}
	\includegraphics[width=170mm]{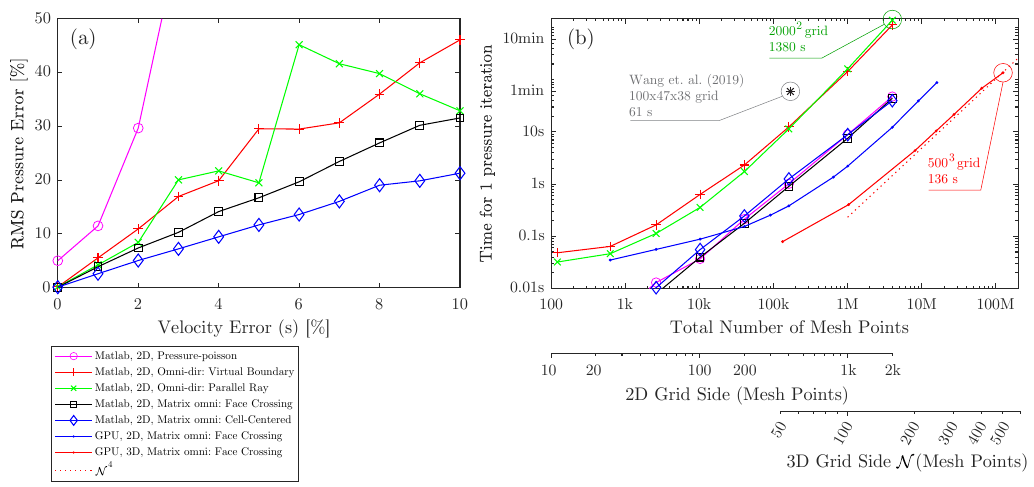}
	\caption{\centering (a) Accuracy of pressure solvers as a function of noise introduced in the velocity field. (b) Time required for 1 iteration of the algorithms considered, including the GPU implementation.}
	\label{fig:err-taylor}
\end{figure*}

A summary of the RMS pressure errors, as a percentage of the maximum absolute pressure in the field, are presented in Figure \ref{fig:err-taylor} (a) for all solvers considered. It becomes quickly evident that the pressure-Poisson solver presents very poor accuracy, providing incorrect results with an RMS error of $\sim5\%$ of the peak suction even when the underlying velocity field contains no noise. The error of the pressures obtained by the pressure-Poisson solver then shoots up very quickly as a function of velocity error. The two omnidirectional methods (``virtual boundary'' and ``parallel ray'') perform better than the pressure Poisson method, providing a somewhat linear relationship with the velocity error with a zero crossing at approximately zero pressure error. However, we can see the statistics from both methods fluctuate for different velocity errors, due to the limited number of rays employed during the integration. We did not run more cases for converged error statistics for the traditional omnidirectional methods, as they are highly time intensive. The matrix omnidirectional methods proposed in this manuscript, on the other hand, offer the lowest errors; the ``cell-centered'' approach performing slightly better. Furthermore, the statistics from the matrix methods are better converged, as the continuous integrals utilized mimic the limit of infinite rays.

When the computational cost to evaluate pressure from the velocity fields is considered, there are several points that will be further elaborated throughout this manuscript. For this first analysis on the 2D Taylor vortex, we consider the computational time required to compute one iteration of each one of the five methods examined in Figure \ref{fig:err-taylor} (b) for a variety of different grid sizes ranging from $11^2$ to $2001^2$ in 2D. We also include in Figure \ref{fig:err-taylor} (b) the computational times for the GPU implementation in 2D and 3D (for grids of $51^3$ to $501^3$), shown as dot symbols. The only method that requires a single iteration to converge is the pressure-Poisson method, and all other methods will require multiple iterations to converge. Less than 30 iterations were required for the Taylor vortex problem examined in this section for all omnidirectional methods, but further discussion is presented in Section \ref{sec:Convergence} regarding the convergence behavior for more complex cases, where hundreds of iterations may be required to achieve satisfactory convergence. In our numerical experiments, however, the matrix omnidirectional methods always converge in less iterations to a given residual than the traditional omnidirectional methods.

Figure \ref{fig:err-taylor} (b) demonstrates how significant the computational time reduction is for the matrix omnidirectional methods discussed herein, requiring a logarithmic scale to fit the times from the different methods. One iteration of the matrix methods requires slightly more time to solve in Matlab than the pressure-Poisson equation, but the times are of the same order. On the other hand, the traditional omnidirectional integration schemes require about 20 times longer per iteration for a given grid size in 2D for the discretization chosen here. The number of iterations required to convergence of the pressure to a certain criterion is at least the same as the matrix methods proposed herein. 

When we consider our GPU implementation of the 2D matrix omnidirectional solver, we observe a $\sim$ 3-3.5$\times$ speedup in comparison to our Matlab implementation. For 3D grids, we only implemented the matrix omnidirectional solver in the GPU, and curiously we find that the 3D grids solve significantly faster for a given precision in the CG solver for the same number of grid points, averaging about $\sim$ 9$\times$ faster than a 2D grid with the same total number of grid points. We have not explored the underlying reason for this drop in computational complexity, but it is likely related to the number of iterations required for the CG solver to propagate information through the domain, which is of the order of the largest dimension in the domain. However, this effect is particularly exciting because it means that a speedup of approximately 3 orders of magnitude ($\sim$ 500-1000$\times$ faster) can be achieved when compared to 2D grids with the traditional omnidirectional methods, which will enable the tackling of large 3D grids that are becoming highly relevant to the experimental fluids community.

Although a 3D implementation of the traditional omnidirectional integration was not currently examined in this work, an important time complexity scaling is evident by examination of the work presented by \citet{wang_gpu-based_2019}: Using $\mathcal{N}$ to denominate the largest size of the grid between the $x$, $y$ and $z$ directions (assuming a box-shaped domain), the number of rays cast is proportional to the projection area of the domain box analyzed at a given spherical angle (i.e., $\sim \mathcal{N}^2$). The number of operations required to compute each integral is $\sim \mathcal{N}$. Furthermore, the number of spherical angles required to tesselate the enclosing sphere to a given grid spacing is proportional to the area of the enclosing sphere ($\sim \mathcal{N}^2$). Therefore, the number of computations required per iteration must be $t\sim \mathcal{O}(\mathcal{N}^5)$, which is a rather strong scaling. We find in our empirical tests of our 3D GPU implementation of the matrix omnidirectional method presented herein that the time complexity scaling is $t\sim \mathcal{O}(\mathcal{N}^{4})$, as shown in Figure \ref{fig:err-taylor} (b), which is significantly more favorable for larger grids. Note that the total number of points in the 3D grid scales as $\sim \mathcal{N}^{3}$. 

A more direct comparison with the performance described by \citet{wang_gpu-based_2019} can be performed by comparing the performance of the graphics cards used in both studies. The RTX A4000 used in our work is capable of 19.2 TFLOPS at single-precision, whereas the RTX 2080 Ti used by \citet{wang_gpu-based_2019} is capable of 13.45 TFLOPS. We performed our calculations in double-precision, which is reported by \citet{wang_gpu-based_2019} to take approximately 3 minutes per iteration for a $100\times 47\times 38$ grid in their machine. A cutout of the JHU turbulence data described in Section \ref{sec:JHUturbulence} was generated with the same $100\times 47\times 38$ grid size, and one iteration of the omnidirectional matrix ``face-crossing'' scheme takes on average 0.082 s. Therefore, the GPU implementation of our method for 3D grids is estimated to be about 1500$\times$ faster than the state of the art \citep{wang_gpu-based_2019} at this specific grid size and likely even faster for larger grid sizes due to the more favorable time complexity scaling of $t\sim \mathcal{O}(\mathcal{N}^{4})$.

\subsection{JHU Isotropic Turbulence Data (3D)} \label{sec:JHUturbulence}

In this section, the performance of the omnidirectional matrix method will be assessed for a more complex flow field for which the exact values of pressure and velocity are known. The ``forced isotropic turbulence'' direct numerical simulation (DNS) dataset provided by the Johns Hopkins University \citep{li_turbulence_2008} was chosen for this analysis. More specifically, the \textit{``isotropic1024coarse''} database was sampled at the 40$^{th}$ frame (time = 0.2535) and adjacent frames for the computation of the time derivatives. 

Grid resolutions of $50^3$, $100^3$, $200^3$ and $500^3$ were considered in this study. The data cutouts taken are not decimated or downsampled to the specified grid resolutions considered to minimize truncation error in the derivative computations. Instead, the cutouts have the same grid spacing as the original data set and therefore the $50^3$ box is truncated from a corner of the original $1024^3$ set and is four times smaller in physical size than the $200^3$ box. This also enables the demonstration of how the omnidirectional matrix method handles incomplete data boundaries, since none of the cases analyzed comprise of periodic boundaries. As boundary conditions are not defined in the omnidirectional matrix method, this poses no additional implementation challenge.

While sampling the JHU turbulence dataset, we noticed that the momentum equations are not balanced when computing the finite difference gradients of the available velocity/pressure fields in an Eulerian grid, even when including the forcing terms; for accuracy levels from second-order to sixth-order. This slight imbalance appears to be related to the fact the original database was solved with a spectral method. We believe that this effect acts as an unnecessary confounding factor that can contaminate the obtained pressure fields but does not pertain to the accuracy of the pressure solver we propose, but to missing information in the underlying velocity field. Therefore, in order to generate noise-contaminated velocity fields based on the correct underlying pressure field, we perform the following operations: 

\begin{enumerate}
	\item Generate the uncontaminated source field $\vec{f}_0$ based on the true pressure with central finite differences: $\vec{f}_0 = \nabla P_{truth}$;
	\item Compute the source term based on the convective derivative $\vec{f}_u = -\rho (D\vec{u}/Dt)$ based on the velocity fields using second-order accurate central finite differences;
	\item Compute a residual source term $\vec{r}=\vec{f}_0-\vec{f}_u$;
	\item Add noise to the velocity field ($\vec{u}_{noisy}$) and compute a noise-contaminated source term $\vec{f}_{u,n} = -\rho (D\vec{u}_{noisy}/Dt)$;
	\item Correct the source term with the residual: $\vec{f}_{0,n}=\vec{f}_{u,n} + \vec{r}$;
	\item Integrate the noise-contaminated source field $\vec{f}_{0,n}$ to obtain pressure.
\end{enumerate}

\begin{figure*} [h!]
	\centering
	\includegraphics[width=150mm]{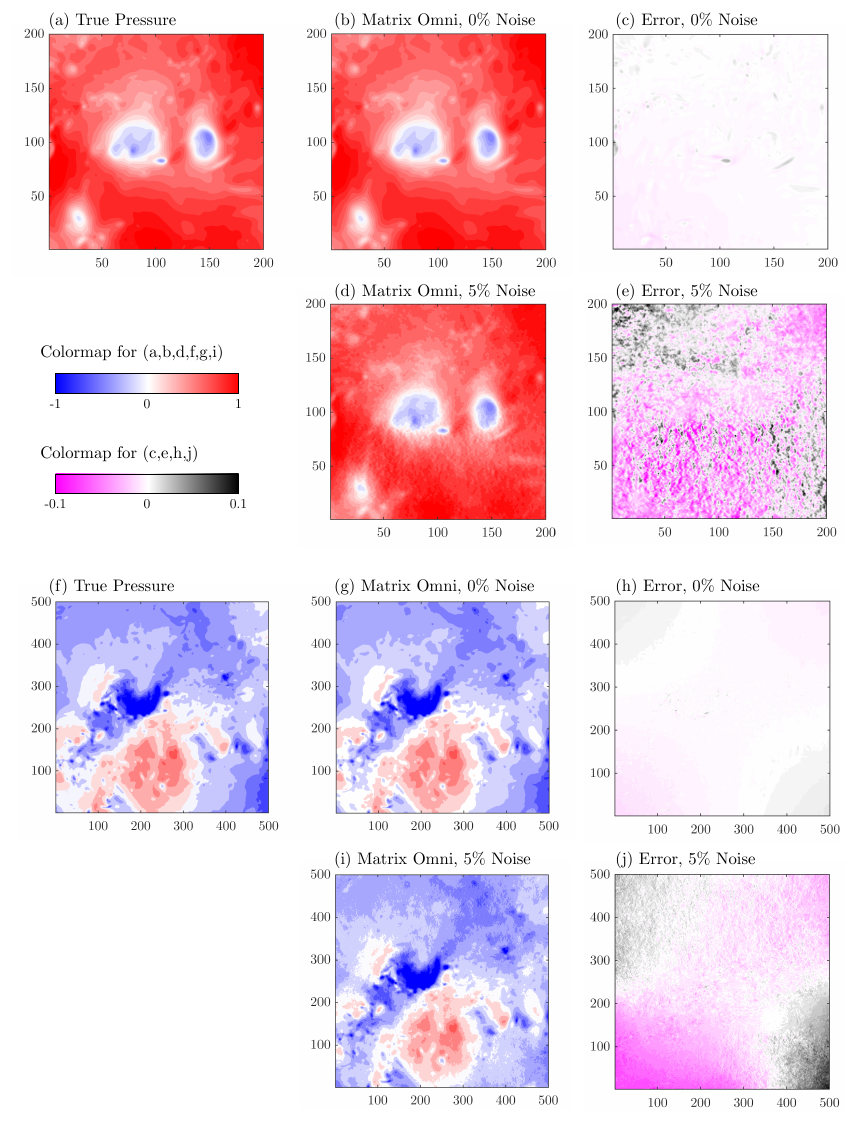}
	\caption{\centering Slice showing pressure contours obtained with the matrix omnidirectional method (``face-crossing'' scheme) for the JHU turbulence database for the $200^3$ box case (a-e) and $500^3$ box case (f-j). 5\% noise corresponds to 1 standard deviation in the raw velocity vectors.}
	\label{fig:JHU-Contours}
\end{figure*}

With this process, the source field $\vec{f}_{0,n}$ computed from the velocity field contaminated with noise of zero magnitude matches $\nabla P_{truth}$ and has no error related to the approximated derivatives. The noise-contaminated velocity fields $\vec{u}_{noisy}$ are generated differently than the previous section, as the fields are three dimensional. The equation below adds multiplicative noise to the velocity fields scaled by a factor $s$, between 0 and 0.1 (0\% and 10\% noise intensity). Note the noise added in Equation \ref{eq:NoisyVectors} is slightly different than Equation \ref{eq:NoisyVectorsTheta} in that it only rescales the vectors without adding a component perpendicular to them. Due to the usage of a single field to estimate pressure, no noise reduction scheme (such as the POD-denoising used in the previous section) was employed. 

\begin{equation} \label{eq:NoisyVectors}
	\vec{u}_{noisy} = \vec{u} (1+s\vec{n})
\end{equation}

Figure \ref{fig:JHU-Contours} displays the contours of pressure for two turbulence box sizes, $200^3$ and $500^3$, which were iterated to a $10^{-3}$ momentum residual as defined in Equation \ref{eq:ResidualMomentum} and further discussed in Section \ref{sec:Convergence}. In each set of cases, the slice is taken at the middle of the respective domain. We can compare the true pressure in Figure \ref{fig:JHU-Contours} (a, f) against the results obtained with the matrix omnidirectional method at $0\%$ noise (b, g) and $5\%$ noise (d, i). The error fields (truth minus ``matrix-omni'' result) are shown in Figure \ref{fig:JHU-Contours} (c, e, h, j) in a tighter color scale. It is evident from the contours in Figure \ref{fig:JHU-Contours} that the matrix omnidirectional results closely approximate the true pressure fields in both cases even with the presence of fairly large velocity noise. A velocity field noise with a $5\%$ standard deviation is representative of the quality that might be achieved with particle image velocimetry for the more challenging experiments where illumination intensity is very low. Nevertheless, we still observe the shape and magnitude of the contours in the noisy cases is close to the true pressure, up to quite significant detail.

\begin{figure*} [h]
	\centering
	\includegraphics[width=100mm]{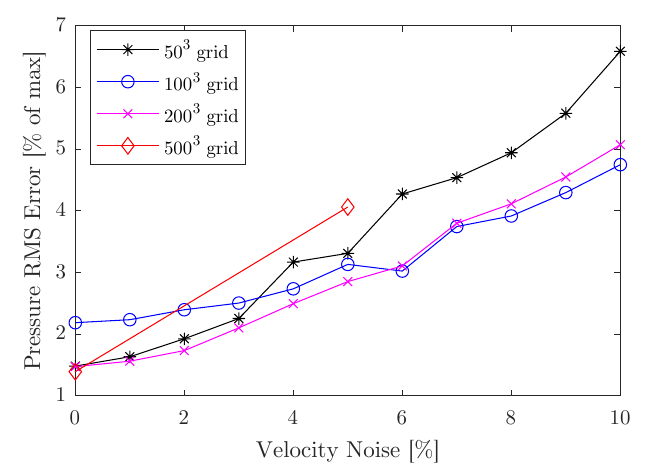}
	\caption{\centering Effect of velocity noise on pressure RMS error for the 3D JHU turbulence data set for the GPU implementation of the matrix omnidirectional ``face-crossing'' scheme.}
	\label{fig:JHU-NoiseIntensity}
\end{figure*}

Several cases with varying velocity noise were tested with this data set and the RMS pressure error is plotted in Figure \ref{fig:JHU-NoiseIntensity} as a function of velocity noise for various grid sizes. We note a similar linear trend as a function of velocity noise as observed in the 2D Taylor vortex shown in Section \ref{sec:TaylorVortex}, however the RMS pressure error is smaller in the JHU turbulence case. This may be related to the more uniformly distributed magnitudes of the velocity vectors in the JHU database, which reduces the error of the velocity gradients required to compute the source terms $\vec{f}$. The generally low error achieved for different grid sizes demonstrates the ability of the matrix omnidirectional method proposed herein to provide accurate pressure estimates from PIV velocity fields even for very large 3D mesh sizes.

\subsection{Limit of Infinite Rays}

In previous sections, we argued that the matrix omnidirectional method we propose simply takes the ``rotating parallel ray'' approach to the limit of infinite rays. From the integral formulation proposed in Section \ref{sec:Concept}, we define the matrix equation weights considering the infinite number of rays coming from all possible angles $\theta$ between 0 and $2\pi$ and all possible ray shifts crossing any given cell in the domain. Therefore, the results from the computations using the ``rotating parallel ray'' approach with increasingly smaller spacing between rays and increasingly more angles should approach the result from the matrix method proposed herein.

\begin{figure*} [h]
	\centering
	\includegraphics[width=100mm]{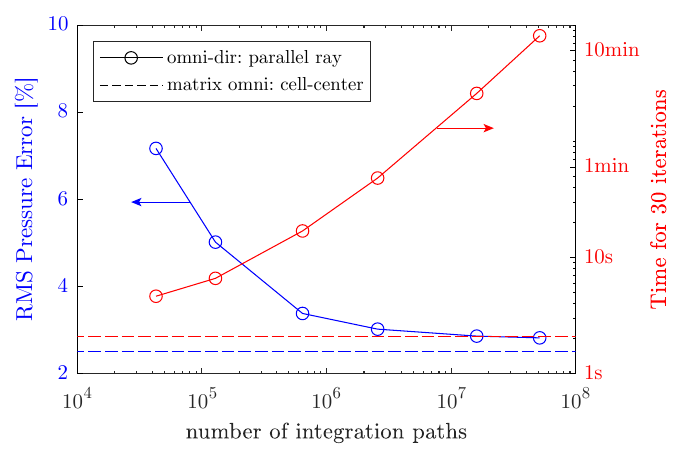}
	\caption{\centering Taking the ``rotating parallel ray'' approach to infinite rays approaches the matrix omnidirectional result.}
	\label{fig:InfiniteRays}
\end{figure*}

This behavior is evidenced in Figure \ref{fig:InfiniteRays}, where the ``rotating parallel ray'' algorithm is run on the 2D Taylor vortex data with 1\% velocity error on a $101^2$ grid for 30 iterations of the Matlab implementation. The blue solid line, showing the RMS pressure error as a percentage of the peak suction at the core of the vortex, evidences this convergence behavior as the number of rays (integration paths) is increased. However, note the logarithmic x-scale, demonstrating a large number of integration paths is required to really achieve this convergence. Evidently, this comes at a high computational cost, which is evidenced by the red solid line in Figure \ref{fig:InfiniteRays}. Compared to the time required to run the 30 iterations of the matrix omnidirectional method (shown as a red dashed line, $\sim 2.07$ s), the cost to run the ``rotating parallel ray'' algorithm to the same accuracy as the matrix omnidirectional method is very high, of order 10 minutes, even for a small $101^2$ domain.

\subsection{Method Convergence} \label{sec:Convergence}

As discussed in Section \ref{MatrixOmni}, Equation \ref{eq:MatrixEquation} is an iterative update equation for the pressure field. It updates both boundary nodes and internal nodes for the $(n+1)^{th}$ iteration. Here we discuss the observed convergence properties of this update method.

In order to define convergence, two metrics were examined. The pressure update metric is defined as:

\begin{equation} \label{eq:ResidualPressure}
	\varepsilon_P^{n+1}=\frac{||\{P^{n+1}\} - \{P^{n}\}||_2}{||\{P^{n+1}\}||_2}
\end{equation}

\noindent where $||\cdot||_2$ is the L2 norm of the field considered. Alternatively, the residual of Equation \ref{eq:MatrixEquation} (momentum residual) can be used as a convergence metric:

\begin{equation}  \label{eq:ResidualMomentum}
	\varepsilon_R^{n+1}=\frac{||[W^{n+1}] \{P^{n+1}\} - [W^{n}] \{P^{n+1}\} -\{S\}||_2}{||\{S\}||_2}
\end{equation}

\noindent where the denominator using only the source term assumes the first residual of Equation \ref{eq:MatrixEquation} was computed with a zero initial guess for $\{P^{n=0}\}=0$. The momentum residual of Equation \ref{eq:ResidualMomentum} is readily available in the CG solver implementation, as it needs to be calculated when the CG solver is initialized.

Although both residuals could be used for assessing convergence, the momentum residual $\varepsilon_R$ was found to be a slightly more conservative convergence criterion (i.e., its decay rate is the same as $\varepsilon_P$ but its magnitude is slightly larger) and therefore it will be the quantity used for all discussions through this manuscript.

Due to the high computational cost of the traditional omnidirectional integration method, previous references typically use a small number of iterations of the algorithm. For example, \citet{McClure2017} used 5 iterations in their examination of a 2D slice of a DNS cylinder flow database, whereas \citet{Liu2006} claim to have observed convergence in less than 10 iterations in a 2D PIV measurement of a cavity shear flow. In the 3D implementation of the omnidirectional integration method by \citet{wang_gpu-based_2019}, only 3-4 iterations were used, which was deemed to have reached convergence by the authors. With the implementation proposed herein, the effect of the number of iterations on convergence can be examined for a very large number of iterations, as the computational cost per iteration is small even for large grids. 

\hfill  \newline

\subsubsection{Effect of Pressure Magnitude at the Boundaries}
\hfill  \newline

First, we examine the convergence behavior for the 2D decaying Taylor vortex at a specified time ($t/T_0=0.1$) and varying the magnitude of the true pressure values at the boundary by changing the extents of the domain considered under the same grid resolution of $101^2$ and no noise contamination of the underlying source fields. 

\begin{figure*} [h]
	\centering
	\includegraphics[width=165mm]{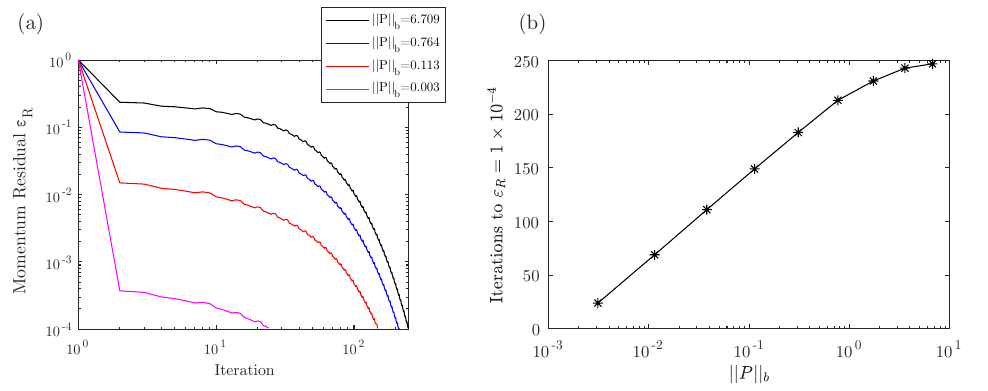}
	\caption{\centering (a) Convergence of the residual as a function of pressure iterations for the 2D Taylor vortex at different domain crop factors. $||P||_b$ is the norm of the pressures at the boundaries. (b) Iterations to a fixed convergence criterion as a function of $||P||_b$.}
	\label{fig:convergence-taylor}
\end{figure*}

The ``pressure magnitude'' here is defined as the norm of the true pressures at the boundary $||P||_b$. It stands to reason that the further the final pressures at the boundary are from the initial guess, the longer it will take for the algorithm to update the boundary values to within a given residual. As the initialization of the pressure term $\{P^{n=0}\}=0$ has zero norm, this effect can be captured by the norm of the boundary pressures $||P||_b$, which is evidenced in Figure \ref{fig:convergence-taylor}. If the domain sampled from the Taylor vortex is very large, then the pressures at the boundary are close to zero (i.e., their norm $||P||_b\rightarrow 0$). Conversely, with a smaller windowing of the domain, the pressures at the boundary will have values further from zero, and their norm grows larger. As evidenced in Figure \ref{fig:convergence-taylor} (a), increasing the norm of the target boundary pressure slows down the convergence of the algorithm. The trend is clearer when the convergence criterion is fixed, as shown in Figure \ref{fig:convergence-taylor} (b), where $\varepsilon_R$ was fixed to $10^{-4}$ and the number of iterations to convergence was counted for several values of $||P||_b$. The number of iterations to convergence appears to be a logarithmic function of the boundary norm $||P||_b$.

This effect may explain why some of the former references observed rapid convergence in the traditional omnidirectional integration method. If the pressure field boundaries include a large region with pressures close to the ambient (or ``far field'') pressure, then a fast convergence rate is attained because the boundary values do not need to be significantly updated. This also serves as a guideline for experimentation; i.e., the convergence rate will be faster when the far field is included in the velocity field measurements. However, this is not possible when the domain includes walls where a pressure distribution is present, as will be demonstrated with real data in Section \ref{sec:Applications}.

\subsubsection{Grid Size}
\hfill  \newline

\begin{figure*} [h]
	\centering
	\includegraphics[width=170mm]{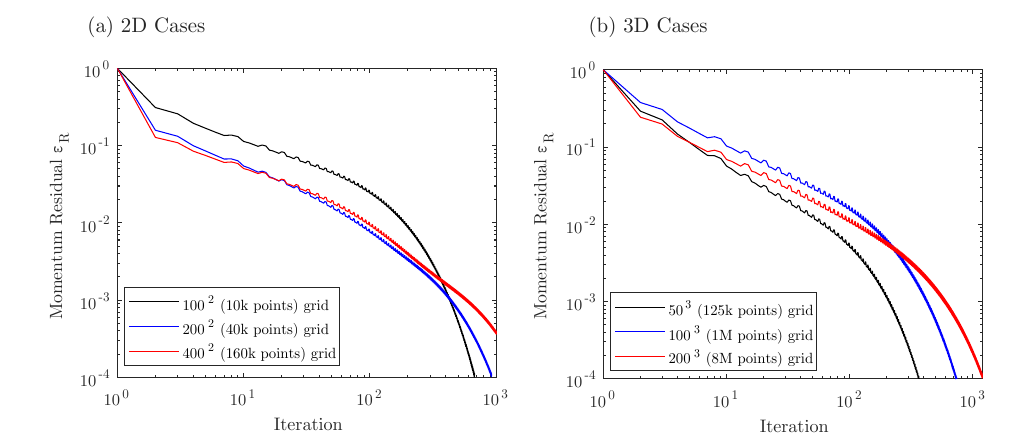}
	\caption{\centering Convergence of the residual as a function of grid size for (a) 2D slices and (b) 3D subvolumes of the JHU isotropic turbulence data set.}
	\label{fig:convergence-gridsize}
\end{figure*}

Now considering fields with complex pressure boundaries, we evaluate the effect of grid size on the number of iterations required to attain convergence. In this section, the same JHU isotropic turbulence data set examined in Section \ref{sec:JHUturbulence} is considered with different grid sizes. The convergence behavior is summarized in Figure \ref{fig:convergence-gridsize}. In the 2D cases shown in Figure \ref{fig:convergence-gridsize} (a), the 2D slices are taken from the center of the 3D box of same size. It is evidenced in both Figure \ref{fig:convergence-gridsize} (a) and (b) that the convergence behavior is somewhat complex with an initial polynomial phase and a rapid exponential convergence at the latter iterations. The ``knee'' of the curves where the exponential phase begins seems to be different for each case analyzed, resulting in crossing of the convergence lines from each case. Generally, however, the convergence is faster for the smaller grids at the smaller residuals $\varepsilon_R$ and less than $10^3$ iterations is required for most cases considered to converge to $\varepsilon_R<10^{-4}$.

Curiously, the number of iterations required to compute the 3D cases (which include far more boundary points) is comparable to the 2D cases. The wall clock time required for the 2D cases, however, is far smaller; taking 310 seconds to run 1,000 iterations for the $400^2$ grid and 2863 seconds to run 1,000 iterations for the $200^3$ grid using the GPU implementation of this algorithm.

\subsubsection{Relationship Between Residual and Error}
\hfill  \newline

When the source field $\{S\}$ used as an input to the omnidirectional matrix algorithm is contaminated by noise, it may be unnecessary to fully converge the pressure integration. Therefore, it is worthwhile to examine the relationship between the residual convergence and the error in the pressure fields for noisy cases to better understand where satisfactory convergence is attained.

\begin{figure*} [h]
	\centering
	\includegraphics[width=165mm]{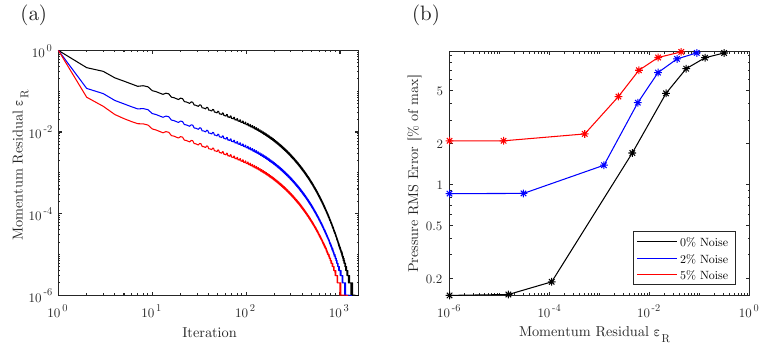}
	\caption{\centering (a) Convergence of the residual as a function of pressure iterations. (b) Effect on error against ground truth for a $100^3$ grid from the JHU dataset.}
	\label{fig:convergence-residual}
\end{figure*}

Figure \ref{fig:convergence-residual} (a) presents the residual convergence for the $100^3$ grid case from the JHU dataset for three different noise magnitude cases: 0\% noise, 2\% noise and 5\% noise. The first observation is that the cases with larger noise converge to lower residuals $\varepsilon_R$ earlier than the uncontaminated case. Figure \ref{fig:convergence-residual} (b) demonstrates that the addition of noise in the underlying velocity fields makes the more converged cases plateau to a given RMS error in pressure when compared to the true pressure. For example, although $\varepsilon_R=10^{-5}$ results in an RMS pressure error of 0.15\% for the uncontaminated case, a $\varepsilon_R\sim10^{-4}$ is sufficient to attain the best pressure error of $\sim 1\%$ for the 2\% noise case and $\varepsilon_R\sim10^{-3}$ is sufficient to attain the RMS error of $\sim 2\%$ for the 5\% noise case. Note that under these conditions, somewhere between 100 and 500 iterations would already be sufficient for convergence, as shown in Figure \ref{fig:convergence-residual} (a), and therefore a momentum residual of $\varepsilon_R\sim10^{-3}$ to $10^{-4}$ may be an appropriate stopping criterion for this algorithm under practical conditions where noise is present in the underlying velocity fields.

\subsubsection{Wall-clock Time Convergence}
\hfill  \newline

\begin{figure*} [h]
	\centering
	\includegraphics[width=85mm]{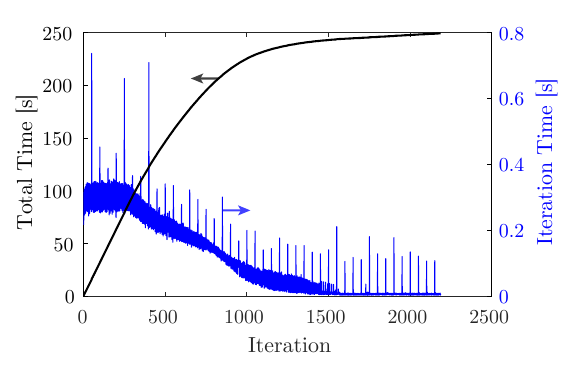}
	\caption{\centering Total wall clock time and time per iteration for the GPU implementation to solve the $100^3$ grid.}
	\label{fig:wallTime}
\end{figure*}

It is also worth mentioning that the use of the Conjugate Gradient algorithm to solve the matrix equation, as implemented in the GPU version of this algorithm, presents the added advantage of enabling the use of the previous iteration's pressure field as a first guess for the CG algorithm. This significantly speeds up the solution time at the latter iterations of the algorithm, as is demonstrated in Figure \ref{fig:wallTime} for the $100^3$ grid from the JHU isotropic turbulence dataset. As it can be seen, after iteration $\sim300$ the time per iteration slowly decreases, reaching a negligible baseline of $\sim 6$ milliseconds after iteration 1500 (corresponding to the various kernel launch overheads in the GPU). The peaks in the iteration times (blue lines) are due to periodic saving of the intermediary fields every 50 iterations.

\section{Application Examples} \label{sec:Applications}

In this section, a few example datasets from previous works by the authors are explored to demonstrate the ability of this technique to accurately reconstruct pressure from PIV measurements without \textit{a priori} knowledge of the pressures at the boundaries, by only using the free stream pressure as the integration constant.

\subsection{Planar PIV, Slanted cylinder flow, center plane slice} \label{sec:PIV_LoftedCyl}

First, we consider the bluff body wake of a cylinder with a slanted afterbody examined by \citet{Zigunov2020} at the slant angle of $\phi=45^\circ$ and diameter-based Reynolds number of $Re_D=25,000$. At these conditions, this flow presents an interesting hysteresis behavior, where the same Reynolds number supports two distinct wake flow topologies. In the ``vortex state'', the flow separates at the leading edge of the slanted surface and reattaches on the slanted surface further downstream, forming a separation bubble and a counter-rotating vortex pair. In the ``wake state'', the flow separates at the same point but does not reattach, forming a more traditional fully-separated bluff body wake. The ``wake state'' is the default state for this wake at low Reynolds numbers, but if the Reynolds number exceeds a threshold value (of about 33,000), then the ``vortex state'' forms and persists if the free stream velocity is then reduced \citep{Zigunov2022}, allowing a range of Reynolds numbers to display two different flow topologies at the same $Re_D$.

The surface pressure distribution was measured using an Omega PX653 pressure transducer with a range of 0-0.1 inH2O, resulting in a $C_p$ uncertainty of $\pm$0.03 for the conditions analyzed. Five hundred planar PIV velocity fields were acquired at the center plane (streamwise-vertical) at a rate of 10 Hz, enabling the approximation of the Reynolds stresses required to compute the average pressure field. 

The average pressure field is calculated by using the Reynolds-Averaged Navier Stokes equations, dropping the viscous terms:

\begin{equation} \label{RANS_f}
	 f_i=\delta_{ij}\frac{\partial \bar{P}}{\partial x_j} = -\rho \bigg(\bar{u}_j\frac{\partial \bar{u}_i}{\partial x_j}  + \frac{\partial (\overline{u_i^\prime u_j^\prime})}{\partial x_j}\bigg)
\end{equation}

\begin{figure*} [h]
	\centering
	\includegraphics[width=165mm]{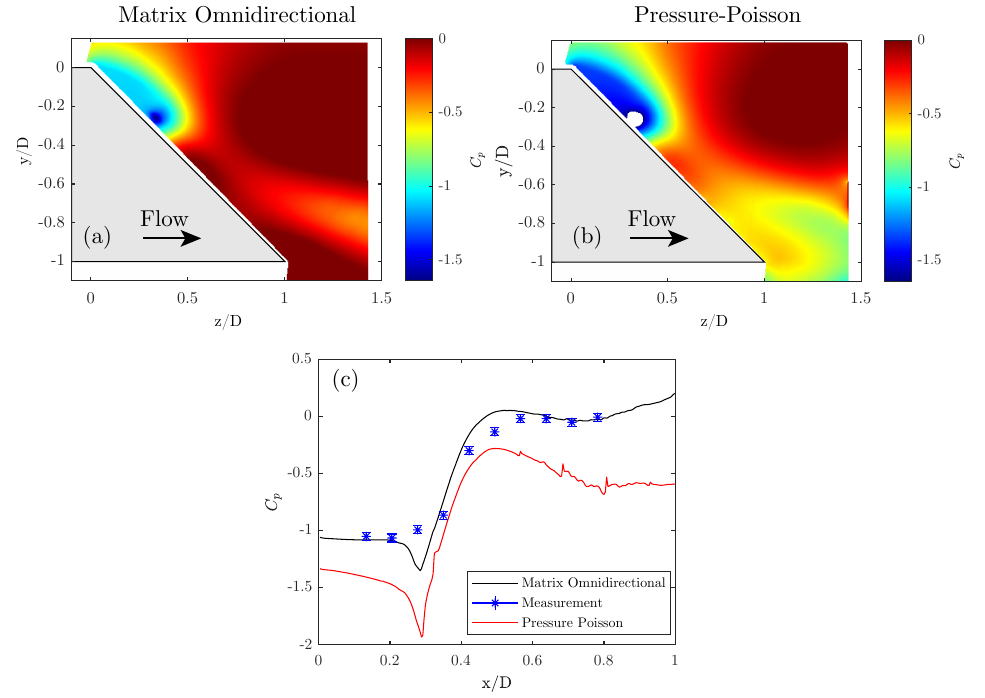}
	\caption{\centering (a,b) Contours of center-plane pressure obtained from PIV measurements for the slanted cylinder model for the ``vortex regime'' using the matrix omnidirectional method (a) and pressure-Poisson method (b). (c) Comparison with surface measurements with traditional transducers.}
	\label{fig:lofted-cylinder}
\end{figure*}

Since the underlying velocity fields were from planar PIV, the out-of-plane derivatives and velocity components were considered equal to zero. As the plane chosen is the symmetry plane of the flow, this is true for the mean quantities but not for the fluctuations.

The forcing vector field $f_i$ is computed using Equation \ref{RANS_f} and integrated using the matrix omnidirectional method using the ``face-crossing'' scheme with our GPU implementation. The PIV vector fields are 427$\times$360 ($\sim$154k grid points) in size and masks were applied to remove regions without valid vectors ($\sim$87k grid points contain valid data). The first iterations took approximately 0.4 seconds, but about 3,000 iterations were required to converge the boundaries; taking about 16 minutes to converge the pressure field with the matrix method. 

Although the number of iterations and convergence time was large for this case, the results presented in Figure \ref{fig:lofted-cylinder} for the ``vortex regime'' demonstrate the motivation for attaining convergence of the boundaries in the matrix omnidirectional method. The contours of pressure obtained with the matrix omnidirectional method in Figure \ref{fig:lofted-cylinder} (a) are clearly a different field than the results from the pressure-Poisson method in Figure \ref{fig:lofted-cylinder} (b), which was obtained by setting the boundary conditions to Neumann boundaries $(\nabla P \cdot \hat{n}=\vec{f}\cdot \hat{n})$. In both cases, the pressure at the top-right corner of the field, which is the closest to a free stream value, is set to $C_p=0$. 

\begin{figure*} [h]
	\centering
	\includegraphics[width=165mm]{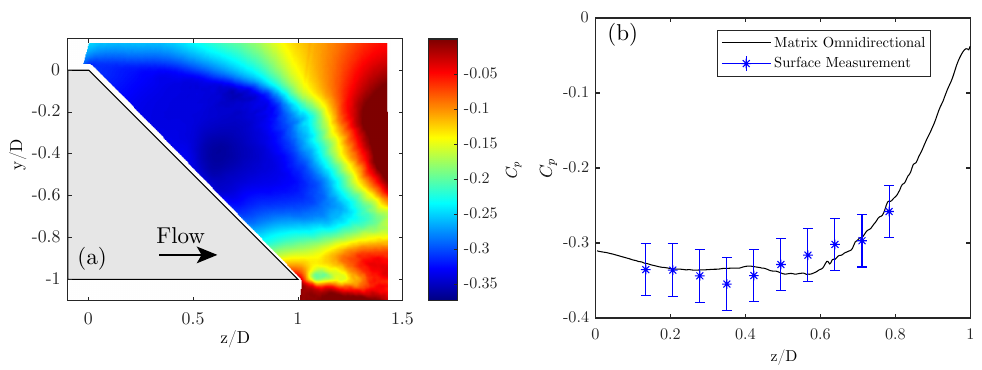}
	\caption{\centering (a) Contours of center-plane pressure obtained for ``wake regime'' for the matrix omnidirectional method. (b) Comparison of pressures at the surface.}
	\label{fig:lofted-poisson}
\end{figure*}

In the comparison with the transducer measurements shown in Figure \ref{fig:lofted-cylinder} (c), it becomes clear that the matrix omnidirectional method presents a better match without any \textit{a priori} knowledge (i.e., the only information provided was the $C_p=0$ value at the top-right corner after pressure was integrated). The magnitude of the surface pressures is off by approximately $\Delta C_p=0.3-0.5$ in the pressure-Poisson method. Furthermore, the low-pressure region towards the trailing edge $z/D\approx 1$ has incorrect topology in the pressure-Poisson solution of Figure \ref{fig:lofted-poisson} (b), incorrectly assigning a suction zone to the trailing edge flow. These aspects demonstrate the limitations of using the pressure-Poisson equation to obtain pressure estimates from PIV fields. Although the pressure-Poisson solution may seem adequate at a first glance without any other reference pressure field for comparison, it may lead to misleading interpretations about the physics of a problem. Therefore, the computational time investment for solving the omnidirectional matrix equation through a large number of iterations is likely justified if more accurate and robust pressure estimates are to be extracted from experimental PIV fields.

For the sake of completeness, the pressure field for the ``wake regime'' of this problem is also provided in Figure \ref{fig:lofted-poisson} (a), only for the matrix omnidirectional method. Note the remarkable agreement with the surface measurements  in Figure \ref{fig:lofted-poisson} (b), likely due to the lower surface-normal pressure gradients. In other words, the pressure is not varying as much away from the surface as in the ``vortex regime'', minimizing the effect of the missing vectors near the surface (due to laser reflections) and enabling a fairer comparison between the surface pressure from transducers and near-surface pressure from PIV. Note the pressure field topology in Figure \ref{fig:lofted-poisson} (a) matches the expected distribution for a fully separated bluff body wake, with the entirety of the wake presenting a moderate suction and the pressure recovering to free stream values towards the trailing edge.

\subsection{Scanning SPIV, Supersonic Double-Fin Flow} \label{sec:DF_SPIV}

In this application we demonstrate the possibility of applying this technique to obtain a 3D average pressure field for a compressible flow at $M_\infty=2$ over a double-fin. The PIV fields are obtained using the newly-proposed ``scanning stereoscopic PIV'' (SSPIV) technique \citep{Zigunov2023}. This large volume ($53\times 66\times 100$ mm) with the $<$1$\mu$m particles required for this supersonic flow is currently outside of the typical range for a tomographic PIV setup, which motivates the choice of this technique. In brief, the SSPIV technique comprises of a continuously scanning stereoscopic PIV system where the entire acquisition optics are slowly scanned throughout the interrogation volume. A stack of about 800 planes is then acquired over the course of a run, and each plane is assigned to a $z$ location according to the timing of the laser Q-switch in relationship to a simultaneously-measured linear potentiometer attached to the traverse mechanism. This allows for the calculation of the average velocities and approximation of the Reynolds stresses by performing a spatio-temporal average, where PIV planes from $z$ locations within a threshold ($\pm$ 1 mm) of a given $z$ coordinate are averaged. Further details about the acquisition technique are provided in \citet{Zigunov2023}, and further details on the flow and analysis of the 3D flow structures observed can be found in \cite{Seckin2023}. As a final note, we point out that the grid spacing is different in the in-plane directions and the out-of-plane direction, which makes the computation of the rectangular grid coefficients provided in Section \ref{sec:FaceCrossing3D} particularly useful.

Before discussing the implementation details, we note that 30 planes are used for each spatiotemporal average plane, and to attain better statistical convergence of the averages and fluctuations the vectors on a $5\times 5\times 30$ grid centered at every grid point are used. More planes could not be acquired due to the limited blowdown time of the facility. This number of vectors is sufficient to converge the mean quantities, as discussed by \citet{Zigunov2023}, but the Reynolds stresses are likely not converged. The contribution of the Reynolds stresses in this flow field is small, and a comparison between the pressure fields with and without including the Reynolds stresses was performed and negligible difference is observed. For this demonstration, the results are reported with the Reynolds stress terms included in the computation of the source field.

To obtain the 3D average pressure field measurements from PIV we use the compressible RANS equations, as described by \citet{vanOudheusden2008}. This requires three key, but reasonable assumptions: (1) adiabatic flow, (2) ideal gas and (3) negligible effect of density fluctuations. The latter assumption (3) is reasonable for moderate Mach numbers. The ideal gas assumption (2) is also valid for this flow, as the minimum compressibility factor is estimated to be $PV/RT=0.998$ \citep{Su1946}. Given assumptions (1) and (2), we can find the temperature field $T$ from the velocity field by knowing the total temperature $T_\infty$: ($T_\infty=173$ K from isentropic relations)

\begin{equation}
	\frac{T}{T_\infty}=1+\frac{\gamma-1}{2} M_\infty^2\bigg(1-\frac{V^2}{V_\infty^2}\bigg)
\end{equation}

\noindent where $\gamma$ is the specific heat ratio and $V$ is the velocity magnitude. Once the temperature field is estimated, the following conservative form of the compressible RANS momentum equations is used: \citep{vanOudheusden2008}

\begin{equation} \label{eq:compressible-pressure}
	\bigg(\delta_{ij}+\frac{\overline{u_i u_j}}{RT}\bigg) \frac{\partial \ln (\bar{P}/P_\infty)}{\partial x_j} = -\frac{1}{RT} \bigg( \frac{\partial \overline{u_i u_j}}{\partial x_j} - \frac{\overline{u_i u_j}}{T}\frac{\partial T}{\partial x_j}\bigg)
\end{equation}

\noindent where $R$ is the ideal gas constant and $\overline{u_i u_j} = \bar{u}_i\bar{u}_j + \overline{u_i^\prime u_j^\prime}$. Equation \ref{eq:compressible-pressure} can be integrated using the matrix omnidirectional method to obtain $\ln (\bar{P}/P_\infty)$, and therefore $\bar{P}/P_\infty$.

\begin{figure*} [h]
	\centering
	\hspace*{-5mm}
	\includegraphics[width=170mm]{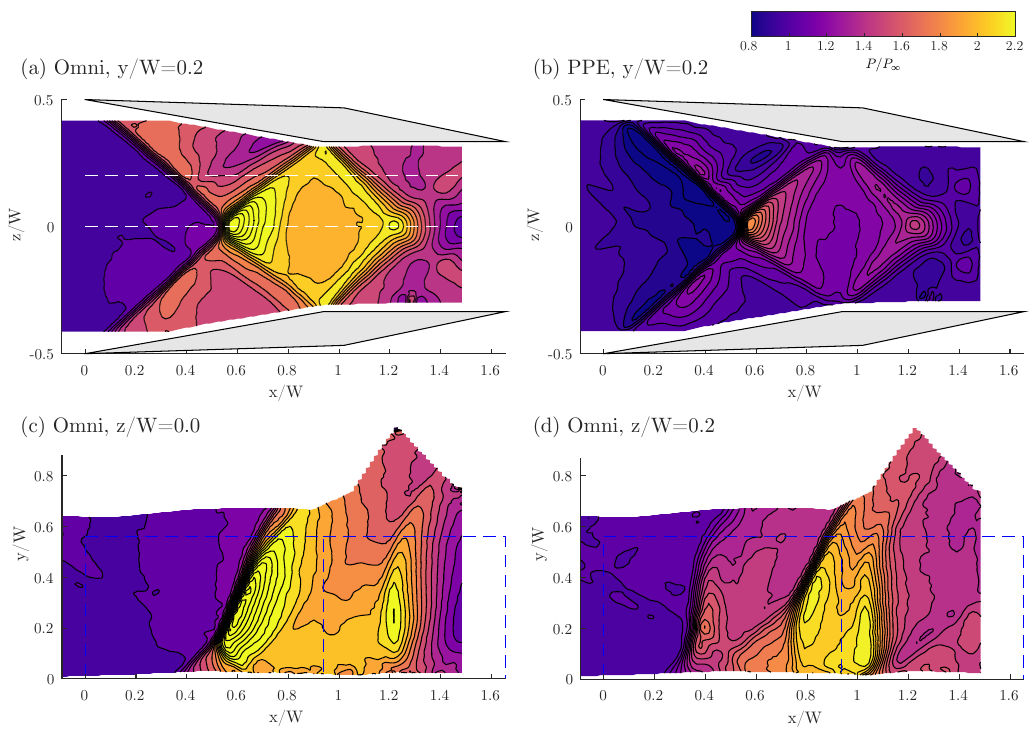}
	\caption{\centering Pressure contours at different slices for the (a, c, d) matrix omnidirectional solution and comparison with (b) pressure-Poisson with Neumann boundary conditions. Dashed white lines in (a) indicate slice locations for (c, d). Dashed blue line in (c, d) indicates fin location. Flow is left to right.}
	\label{fig:dual-fin-pressure-3d}
\end{figure*}

A few slices of the pressure field obtained with the matrix omnidirectional method, comprising of $212\times 266\times 101 = 5.6$ million grid points, are presented in Figure \ref{fig:dual-fin-pressure-3d} (a, c, d). The spatial coordinates are normalized with respect to the fin spacing $W=63.5$ mm. A comparison slice at $y/W=0.2$ showing the pressure-Poisson solution with Neumann boundary conditions is also provided in Figure \ref{fig:dual-fin-pressure-3d} (b), showing how far off the Poisson equation solution is, even though the same reference point at $\{x/W, y/W, z/W\}=\{0,0.2,0\}$ was used in both cases, being enforced to $P/P_\infty=1$.

\begin{figure*} [h]
	\centering
	\hspace*{-5mm}
	\includegraphics[width=170mm]{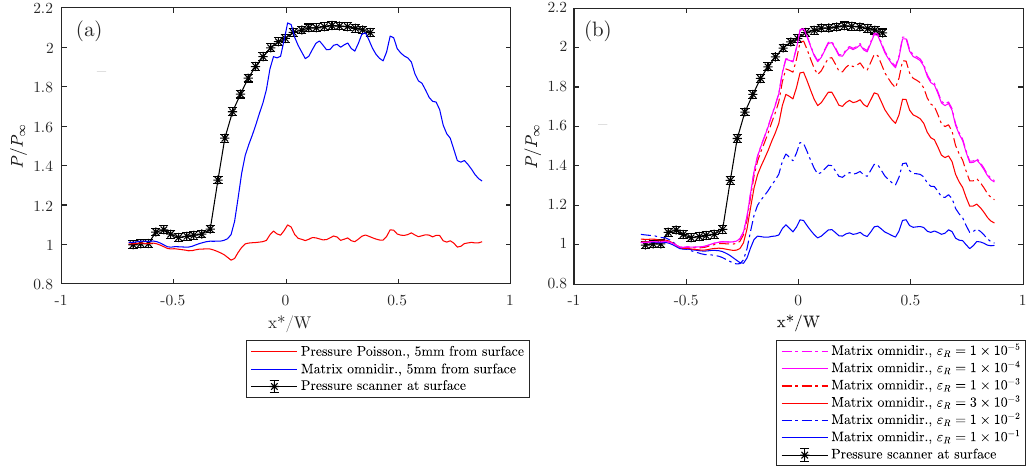}
	\caption{\centering Comparison of the pressures obtained from PIV measurements with pressure scanner pressures. (a) Comparing pressure-Poisson and matrix omnidirectional. (b) Convergence behavior for the matrix omnidirectional method.}
	\label{fig:dual-fin-pressure-comparison}
\end{figure*}

Before further discussion is provided on the pressure fields shown in Figure \ref{fig:dual-fin-pressure-3d} (a, c, d) is provided, we first demonstrate that the pressure fields obtained by the omnidirectional matrix method are a good representation of the true pressures. In Figure \ref{fig:dual-fin-pressure-comparison} (a) the pressures obtained by solving the omnidirectional matrix equation are presented at the center-plane, at the closest positions from the surface of the model ($\sim$ 5 mm away from the surface), and compared with the pressures measured at the surface by a set of pressure taps connected to a Kulite KMPS-1-32-10psid pressure scanner (details in \citet{Seckin2023a}). The pressures from the omnidirectional matrix method (blue line) match very well with the measured surface pressures, being spatially displaced towards the downstream direction as they are inside the lambda shock structure and miss some of the boundary layer effects. The pressures from the Poisson solver, however, are not a match with our measurements. Different boundary conditions for the pressure-Poisson solver were tested (only considering the information known a \textit{priori}, i.e., the free stream pressure), none of which yielded any better results. In Figure \ref{fig:dual-fin-pressure-comparison} (b), a comparison of different residual cutoffs is performed for the dual fin case for the omnidirectional matrix method. For the residual $\epsilon_R=10^{-4}$, 1072 iterations were required, taking 27 minutes to converge. The most converged case tested, at  $\epsilon_R=10^{-5}$, took 2013 iterations (40 minutes). The convergence behavior at the boundary is a scaling of the entire pressure field towards the converged value. The same behavior was observed for the slanted cylinder flow presented in Section \ref{sec:PIV_LoftedCyl}. 

Although the time required to converge (27 minutes) is fairly long for one field (the average pressure field), the mesh size is far larger than was previously explored in past studies. The pressure fields displayed in Figure \ref{fig:dual-fin-pressure-3d} (a, c, d) are data that was previously unavailable for this problem coming exclusively from experimental measurements, and highlights the main features of this flow with unprecedented detail. Features such as the oblique fin-tip shocks and reflected shocks are revealed to have a three-dimensionality due to the open top surface of this double-fin geometry. An expansion fan at the more downstream locations ($x/W\sim 1.2-1.4$) is also revealed. Close to the floor surface where the shock interacts with the boundary layer, a large lambda shock is observed at the center plane. Especially for the double-fin problem, these features are difficult to see with a shadowgraph apparatus due to the difficult optical access owing to the presence of the fins.

\section{Conclusion}

In this work, we demonstrate a new approach for performing omnidirectional pressure integration in experimentally-acquired velocity fields. The approach presented considers all possible ray displacements and ray shifts through a double projection integral that computes partial weights for the neighboring cells of a uniformly-spaced rectangular grid. This approach recasts the omnidirectional ray integration method into a matrix inversion problem, which can be solved far more efficiently.

We demonstrate this approach in two synthetic data sets and two experimental data sets, one of which is a compressible flow field with shocks and expansion waves. We find that the error in the reconstructed pressure is significantly lower in noisy velocity fields than the previous approaches described in the literature, including the omnidirectional ray integration this approach is based upon. Furthermore, we find that the computational cost for solving the matrix omnidirectional equation for one iteration is similar to solving the pressure-Poisson equation. However, many iterations are required to converge the boundary pressures for the matrix omnidirectional method. On the other hand, we find that the number of iterations required to converge the boundaries is always lower than the current implementations of the omnidirectional ray integration approach, since our approach takes the limit of ray integration to an infinite number of rays. The computational cost of the matrix omnidirectional approach is $\sim$ 20 times lower in 2D grids and potentially $>1000$ times lower in 3D grids, with a more favorable scaling of $\mathcal{O}(\mathcal{N}^4)$ instead of $\mathcal{O}(\mathcal{N}^5)$. We successfully demonstrate this method on a grid up to $500^3$ (125 million points) in size, which was previously unattainable with ray integration approaches.

In our demonstration of this approach with two experimental data sets, we see that a fairly large number of iterations is required to converge realistic data where some of the boundaries are near walls. However, the converged pressure fields match well with experimentally-measured pressures at selected locations, providing confidence that the fields obtained are an accurate representation of the real pressure fields and opening exciting new possibilities for volumetric pressure estimation in experimental compressible and incompressible flows.

\section*{Appendix: Example of matrix equation construction for a 3x3 domain}
\begin{figure*} [h]
	\centering
	\hspace*{-5mm}
	\includegraphics[width=70mm]{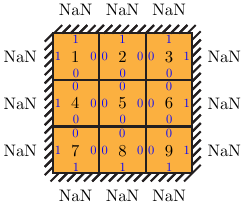}
	 \caption{\centering Example domain for equation construction. NaN's indicate locations where data is not available (i.e., boundaries). Small blue numbers are the value of the boolean $b_j$ for the neighboring cell. Numbers in black at the cell center are the cell index.}
	\label{fig:exampledomain}
\end{figure*}

In this appendix, we show details of how the matrix Equation \ref{eq:MatrixEquation} is constructed for a simple 2D domain consisting of square cells ($\Delta x = \Delta y = \Delta$) of size 3$\times$3 cells using the ``face-crossing'' method depicted in Figure \ref{fig:line-integration} (a). The example domain is depicted in Figure \ref{fig:exampledomain} for reference.

To compute the weights $w$ in the weight matrices $[W]^i$, we need to know the ray counters $A^i_j$ for each of the nine cells in the domain and the total ray count for each cell ($A_{tot}$). From Equation \ref{eq:Atot2D}, we find that $A_{tot}=8\Delta$ for this domain comprising of square cells.

To compute the ray counters for the different boundary types encountered in this domain, we first precompute the constants $c$ using Equation \ref{eq:cxx2D}:

\begin{equation}
	\Delta d = \Delta \sqrt{2}
\end{equation}

\begin{equation}
	\begin{split}
		c_{xx}&=\Delta (2\sqrt{2}-2) \\
		c_{yy}&=\Delta (2\sqrt{2}-2) \\
		c_{xy}&=\Delta (2-\sqrt{2})
	\end{split}
\end{equation}

For a 2D domain in the ``face-crossing'' scheme, there are 10 ray counters $A^i_j$ for each cell. See Equations \ref{eq:wcnp1}, \ref{eq:Acn}, \ref{eq:AEnp1}, \ref{eq:AEn}, \ref{eq:ANnp1} and \ref{eq:ANn} for the equation forms used in their calculations. For the corner cell with index (1) in Figure \ref{fig:exampledomain}, these coefficients are:

\begin{equation}
	\begin{aligned}
		&A_C^{n}=4\Delta \\
		&A_C^{n+1}=0 \\
		&A_E^{n}=c_{xy} \\
		&A_E^{n+1}=c_{xy}+c_{xx} \\
		&A_W^{n}=0 \\
		&A_W^{n+1}=0 \\
		&A_N^{n}=0 \\
		&A_N^{n+1}=0 \\
		&A_S^{n}=c_{xy} \\
		&A_S^{n+1}=c_{xy}+c_{xx} \\
	\end{aligned}
\;\; \textrm{(Cell 1)}
\end{equation}

Note all coefficients sum to $A_{tot}$ (in this case, $8\Delta$), which is always required. The weights $w^i_j$ are then trivially computed from $A^i_j$ using Equation \ref{eq:weights}. For all other cells, we provide the coefficients for this example on Table \ref{tab:Coeffs}.

If a cell within the domain was missing data, the coefficients for the neighboring cells, as well as the neighbor's neighbors, would change. As an example, if cell (1) contained a NaN value, then cells (2), (3), (4), (5) and (7) would have different values for the $A^i_j$ coefficients. All other cells would be unaffected.

The matrix equation following Equation \ref{eq:MatrixEquation} is already too large in this example to be written with all coefficients. However, using Table \ref{tab:Coeffs} and Equation \ref{eq:weights}, we can express the matrix equation for this example as the following pentadiagonal matrix:

\begin{table} [h!]
	\centering
	\scriptsize{
		\begin{tabular}{lccccccccc}
			\hline
			Cell  & 1 & 2 & 3 & 4 & 5 & 6 & 7 & 8 & 9 \\ \hline
			$A_C^{n}$ & $4\Delta$ & $2\Delta$ & $4\Delta$ & $2\Delta$ & $0$ & $2\Delta$ & $4\Delta$ & $2\Delta$ & $4\Delta$ \\
			$A_C^{n+1}$ & $0$ & $0$ & $0$ & $0$ & $0$ & $0$ & $0$ & $0$ & $0$ \\
			$A_E^{n}$ & $c_{xy}$ & $c_{xy}+c_{xx}$ & $0$ & $0$ & $c_{xx}$ & $0$ & $c_{xy}$ & $c_{xy}+c_{xx}$ & $0$ \\
			$A_E^{n+1}$ & $c_{xy}+c_{xx}$ & $c_{xy}$ & $0$ & $2\Delta$ & $2c_{xy}$ & $0$ & $c_{xy}+c_{xx}$ & $c_{xy}$ & $0$ \\
			$A_W^{n}$ & $0$ & $c_{xy}+c_{xx}$ & $c_{xy}$ & $0$ & $c_{xx}$ & $0$ & $0$ & $c_{xy}+c_{xx}$ & $c_{xy}$ \\
			$A_W^{n+1}$ & $0$ & $c_{xy}$ & $c_{xy}+c_{xx}$ & $0$ & $2c_{xy}$ & $2\Delta$ & $0$ & $c_{xy}$ & $c_{xy}+c_{xx}$ \\
			$A_N^{n}$ & $0$ & $0$ & $0$ & $c_{xy}$ & $c_{yy}$ & $c_{xy}+c_{yy}$ & $c_{xy}$ & $0$ & $c_{xy}$ \\
			$A_N^{n+1}$ & $0$ & $0$ & $0$ & $c_{xy}+c_{yy}$ & $2c_{xy}$ & $c_{xy}$ & $c_{xy}+c_{yy}$ & $2\Delta$ & $c_{xy}+c_{yy}$ \\
			$A_S^{n}$ & $c_{xy}$ & $0$ & $c_{xy}$ & $c_{xy}+c_{yy}$ & $c_{yy}$ & $c_{xy}+c_{yy}$ & $0$ & $0$ & $0$ \\
			$A_S^{n+1}$ & $c_{xy}+c_{yy}$ & $2\Delta$ & $c_{xy}+c_{yy}$ & $c_{xy}$ & $2c_{xy}$ & $c_{xy}$ & $0$ & $0$ & $0$ \\ \hline
	\end{tabular}}
	\caption{Coefficients $A^i_j$ for the simple 3$\times$3 grid example.} \label{tab:Coeffs}
\end{table}

\begin{equation} \label{eq:GiantMatrix}
	\scriptsize
	\begin{split}
	\begin{bmatrix}
		w_{C,1}^{n+1}-1 & w_{E,1}^{n+1} & 0 & w_{S,1}^{n+1} & 0 & 0 & 0 & 0 & 0 \\
		w_{W,2}^{n+1} & w_{C,2}^{n+1}-1 & w_{E,2}^{n+1} & 0 & w_{S,2}^{n+1} & 0 & 0 & 0 & 0 \\
		0 & w_{W,3}^{n+1} & w_{C,3}^{n+1}-1 & 0 & 0 & w_{S,3}^{n+1} & 0 & 0 & 0 \\
		w_{N,4}^{n+1} & 0 & 0 & w_{C,4}^{n+1}-1 & w_{E,4}^{n+1} & 0 & w_{S,4}^{n+1} & 0 & 0 \\
		0 & w_{N,5}^{n+1} & 0 & w_{W,5}^{n+1} & w_{C,5}^{n+1}-1 & w_{E,5}^{n+1} & 0 & w_{S,5}^{n+1} & 0 \\
		0 & 0 & w_{N,6}^{n+1} & 0 & w_{W,6}^{n+1} & w_{C,6}^{n+1}-1 & 0 & 0 & w_{S,6}^{n+1} \\
		0 & 0 & 0 & w_{N,7}^{n+1} & 0 & 0 & w_{C,7}^{n+1}-1 & w_{E,7}^{n+1} & 0 \\
		0 & 0 & 0 & 0 & w_{N,8}^{n+1} & 0 & w_{W,8}^{n+1} & w_{C,8}^{n+1}-1 & w_{E,8}^{n+1} \\
		0 & 0 & 0 & 0 & 0 & w_{N,9}^{n+1} & 0 & w_{W,9}^{n+1} & w_{C,9}^{n+1}-1
	\end{bmatrix} \begin{bmatrix}
	P_1^{n+1} \\
	P_2^{n+1} \\
	P_3^{n+1} \\
	P_4^{n+1} \\
	P_5^{n+1} \\
	P_6^{n+1} \\
	P_7^{n+1} \\
	P_8^{n+1} \\
	P_9^{n+1}
\end{bmatrix} = \dots \\
-\begin{bmatrix}
	w_{C,1}^{n} & w_{E,1}^{n} & 0 & w_{S,1}^{n} & 0 & 0 & 0 & 0 & 0 \\
	w_{W,2}^{n} & w_{C,2}^{n} & w_{E,2}^{n} & 0 & w_{S,2}^{n} & 0 & 0 & 0 & 0 \\
	0 & w_{W,3}^{n} & w_{C,3}^{n} & 0 & 0 & w_{S,3}^{n} & 0 & 0 & 0 \\
	w_{N,4}^{n} & 0 & 0 & w_{C,4}^{n} & w_{E,4}^{n} & 0 & w_{S,4}^{n} & 0 & 0 \\
	0 & w_{N,5}^{n} & 0 & w_{W,5}^{n} & w_{C,5}^{n} & w_{E,5}^{n} & 0 & w_{S,5}^{n} & 0 \\
	0 & 0 & w_{N,6}^{n} & 0 & w_{W,6}^{n} & w_{C,6}^{n} & 0 & 0 & w_{S,6}^{n} \\
	0 & 0 & 0 & w_{N,7}^{n} & 0 & 0 & w_{C,7}^{n} & w_{E,7}^{n} & 0 \\
	0 & 0 & 0 & 0 & w_{N,8}^{n} & 0 & w_{W,8}^{n} & w_{C,8}^{n} & w_{E,8}^{n} \\
	0 & 0 & 0 & 0 & 0 & w_{N,9}^{n} & 0 & w_{W,9}^{n} & w_{C,9}^{n}
\end{bmatrix} \begin{bmatrix}
P_1^{n} \\
P_2^{n} \\
P_3^{n} \\
P_4^{n} \\
P_5^{n} \\
P_6^{n} \\
P_7^{n} \\
P_8^{n} \\
P_9^{n}
\end{bmatrix} - \begin{bmatrix}
S_1 \\
S_2 \\
S_3 \\
S_4 \\
S_5 \\
S_6 \\
S_7 \\
S_8 \\
S_9
\end{bmatrix}
\end{split}
\end{equation}

\noindent where $w_{E,5}^{n}$, for example, is the weight corresponding to cell $5$'s eastern neighbor for iteration $n$. Note that the right hand side of Equation \ref{eq:GiantMatrix} can always be calculated, even for an initial guess, and is a vector. Also note that in the 3D case the matrix would be heptadiagonal. The terms in the source vector $\{S\}$ in the right-hand side have the form provided in Equation \ref{eq:ExplicitEqn_FaceCrossing}. Two examples are provided below:

\begin{equation}
	S_1=-\frac{\Delta}{8} [f_x(2)+f_x(1)-f_y(4)-f_y(1)]
\end{equation}

\begin{equation}
	\begin{split}
	S_5=-\frac{\Delta}{8} [f_x(6)+f_x(5)-f_x(4)-f_x(5)+f_y(2)+f_y(5)-f_y(8)-f_y(5)] \\
	S_5=-\frac{\Delta}{8} [f_x(6)-f_x(4)+f_y(2)-f_y(8)]
	\end{split}
\end{equation}

Solving Equation \ref{eq:GiantMatrix} for an initial guess $\{P\}^0$ would provide one iteration of update for $\{P\}^1$. The value of $\{P\}^1$ can then be placed on the right-hand side of Equation \ref{eq:GiantMatrix} to solve for $\{P\}^2$, and so forth. $\{P\}^{n+1}$ appears to converge for this scheme in all tests we have performed, and we have never observed divergence with a starting guess of $\{P\}^0=0$. However, we have not examined nor proved that this is always true for any initial guess, and the proof of convergence for the general case is left as future work.

\section*{Data Availability Statement}

The data that support the findings of this study will be openly available at the following url: \url{https://dx.doi.org/10.5281/zenodo.10697026}.

%\section*{Acknowledgements}
%	This work was supported by the U.S. Department of Energy through the Los Alamos National Laboratory. Los Alamos National Laboratory is operated by Triad National Security, LLC, for the National Nuclear Security Administration of U.S. Department of Energy (Contract No. 89233218CNA000001).

\newcommand{\newblock}{}
\bibliographystyle{apalike}
\bibliography{Pressure-calculations}

\begin{thebibliography}{}

\bibitem[Charonko et~al., 2010]{charonko_assessment_2010}
Charonko, J.~J., King, C.~V., Smith, B.~L., and Vlachos, P.~P. (2010).
\newblock Assessment of pressure field calculations from particle image
  velocimetry measurements.
\newblock {\em Measurement Science and Technology}, 21(10):105401.

\bibitem[Faiella et~al., 2021]{faiella_error_2021}
Faiella, M., Macmillan, C. G.~J., Whitehead, J.~P., and Pan, Z. (2021).
\newblock Error propagation dynamics of velocimetry-based pressure field
  calculations (2): on the error profile.
\newblock {\em Measurement Science and Technology}, 32(8):084005.

\bibitem[Li et~al., 2008]{li_turbulence_2008}
Li, Y., Perlman, E., Wan, M., Yang, Y., Meneveau, C., Burns, R., Chen, S.,
  Szalay, A., and Eyink, G. (2008).
\newblock A public turbulence database cluster and applications to study
  lagrangian evolution of velocity increments in turbulence.
\newblock {\em Journal of Turbulence}, 9:N31.

\bibitem[Liu and Katz, 2006]{Liu2006}
Liu, X. and Katz, J. (2006).
\newblock Instantaneous pressure and material acceleration measurements using a
  four-exposure piv system.
\newblock {\em Experiments in Fluids}, 41(2):227--240.

\bibitem[Liu and Moreto, 2020]{liu_error_2020}
Liu, X. and Moreto, J.~R. (2020).
\newblock Error propagation from the {PIV}-based pressure gradient to the
  integrated pressure by the omnidirectional integration method.
\newblock {\em Measurement Science and Technology}, 31(5):055301.

\bibitem[Liu et~al., 2016]{liu_instantaneous_2016}
Liu, X., Moreto, J.~R., and Siddle-Mitchell, S. (2016).
\newblock Instantaneous {Pressure} {Reconstruction} from {Measured} {Pressure}
  {Gradient} using {Rotating} {Parallel} {Ray} {Method}.
\newblock In {\em 54th {AIAA} {Aerospace} {Sciences} {Meeting}}. American
  Institute of Aeronautics and Astronautics.

\bibitem[McClure and Yarusevych, 2017]{McClure2017}
McClure, J. and Yarusevych, S. (2017).
\newblock Optimization of planar piv-based pressure estimates in laminar and
  turbulent wakes.
\newblock {\em Experiments in Fluids}, 58(5):62.

\bibitem[Nie et~al., 2022]{nie_error_2022}
Nie, M., Whitehead, J.~P., Richards, G., Smith, B.~L., and Pan, Z. (2022).
\newblock Error propagation dynamics of {PIV}-based pressure field calculation
  (3): what is the minimum resolvable pressure in a reconstructed field?
\newblock {\em Experiments in Fluids}, 63(11):168.

\bibitem[Pan et~al., 2016]{pan_error_2016}
Pan, Z., Whitehead, J., Thomson, S., and Truscott, T. (2016).
\newblock Error propagation dynamics of {PIV}-based pressure field
  calculations: {How} well does the pressure {Poisson} solver perform
  inherently?
\newblock {\em Measurement Science and Technology}, 27(8):084012.

\bibitem[Seckin et~al., 2023a]{Seckin2023a}
Seckin, S., Mears, L., Song, M., Alvi, F., and Zigunov, F. (2023a).
\newblock Surface properties of double-fin generated shock-wave/boundary-layer
  interactions.
\newblock {\em AIAA Journal}, 61(12):5302--5319.

\bibitem[Seckin et~al., 2023b]{Seckin2023}
Seckin, S., Song, M., Zigunov, F., and Alvi, F. (2023b).
\newblock Three-dimensional flow characterization of double-fin {SBLI}.
\newblock In {\em AIAA Aviation 2023 Forum}.

\bibitem[Sperotto et~al., 2022]{Sperotto2022}
Sperotto, P., Pieraccini, S., and Mendez, M.~A. (2022).
\newblock A meshless method to compute pressure fields from image velocimetry.
\newblock {\em Measurement Science and Technology}, 33(9):094005.

\bibitem[Su, 1946]{Su1946}
Su, G.-J. (1946).
\newblock Modified law of corresponding states for real gases.
\newblock {\em Industrial {\&} Engineering Chemistry}, 38(8):803--806.

\bibitem[van Oudheusden, 2008]{vanOudheusden2008}
van Oudheusden, B.~W. (2008).
\newblock Principles and application of velocimetry-based planar pressure
  imaging in compressible flows with shocks.
\newblock {\em Experiments in Fluids}, 45(4):657--674.

\bibitem[van Oudheusden, 2013]{van_oudheusden_piv-based_2013}
van Oudheusden, B.~W. (2013).
\newblock {PIV}-based pressure measurement.
\newblock {\em Measurement Science and Technology}, 24(3):032001.

\bibitem[Wang et~al., 2019]{wang_gpu-based_2019}
Wang, J., Zhang, C., and Katz, J. (2019).
\newblock {GPU}-based, parallel-line, omni-directional integration of measured
  pressure gradient field to obtain the {3D} pressure distribution.
\newblock {\em Experiments in Fluids}, 60(4):58.

\bibitem[Zhang et~al., 2022]{Zhang2022}
Zhang, J., Bhattacharya, S., and Vlachos, P.~P. (2022).
\newblock Uncertainty of piv/ptv based eulerian pressure estimation using
  velocity uncertainty.
\newblock {\em Measurement Science and Technology}, 33(6):065303.

\bibitem[Zigunov et~al., 2023]{Zigunov2023}
Zigunov, F., Seckin, S., Huss, R., Eggart, C., and Alvi, F. (2023).
\newblock A continuously scanning spatiotemporal averaging method for obtaining
  volumetric mean flow measurements with stereoscopic {PIV}.
\newblock {\em Experiments in Fluids}, 64(3):56.

\bibitem[Zigunov et~al., 2020]{Zigunov2020}
Zigunov, F., Sellappan, P., and Alvi, F. (2020).
\newblock Reynolds number and slant angle effects on the flow over a slanted
  cylinder afterbody.
\newblock {\em Journal of Fluid Mechanics}, 893:A11.

\bibitem[Zigunov et~al., 2022]{Zigunov2022}
Zigunov, F., Sellappan, P., and Alvi, F. (2022).
\newblock Hysteretic flow regime switching in the wake of a cylinder with a
  slanted afterbody.
\newblock {\em Experiments in Fluids}, 63(5):80.

\end{thebibliography}

\end{document}